\documentclass[times,twocolumn,final,authoryear]{elsarticle}
\usepackage{jasr}
\usepackage{framed,multirow}

\usepackage{amssymb}
\usepackage{latexsym}

\usepackage[switch]{lineno}

\usepackage{url}
\usepackage{xcolor}
\definecolor{newcolor}{rgb}{.8,.349,.1}
\usepackage[normalem]{ulem}

\usepackage[citebordercolor=white]{hyperref}

\journal{Advances in Space Research}

\begin{document}

\verso{Samuel Skirvin \textit{etal}}

\begin{frontmatter}

\title{Small-scale solar jet formation and their associated waves and instabilities} %\tnoteref{tnote1}}

%\tnotetext[tnote1]{This is an example for title footnote coding.}

\author[1]{Samuel \snm{Skirvin}\corref{cor1}}
\cortext[cor1]{Corresponding author: \ead{samuel.skirvin@kuleuven.be}}
\author[3]{Gary \snm{Verth}}  %\fnref{fn1}}
%\fntext[fn1]{This is author footnote for second author.}
\author[4]{Jos\'e Juan \snm{Gonz\'alez-Avil\'es}}
\author[5]{Sergiy \snm{Shelyag}}
\author[6]{Rahul \snm{Sharma}}
\author[7]{Francisco S. \snm{Guzm\'an}}
\author[3]{Istvan \snm{Ballai}}
\author[6]{Eamon \snm{Scullion}}
\author[2]{Suzana \snm{S. A. Silva}}
\author[2]{Viktor \snm{Fedun}}

\address[1]{Center for mathematical Plasma Astrophysics, Mathematics Department, KU Leuven, Celestijnenlaan 200B bus 2400, B-3001, Leuven, Belgium}
\address[2]{Plasma Dynamics Group, Department of Automatic Control and Systems Engineering, The University of Sheffield, Mappin Street, Sheffield S1 3JD, UK}
\address[3]{Plasma Dynamics Group, School of Mathematics and Statistics, The University of Sheffield, Hicks Building, Hounsfield Road, Sheffield S3 7RH, UK}
\address[4]{Investigadores por M\'exico-CONACYT, Servicio de Clima Espacial M\'exio, Laboratorio Nacional de Clima Espacial, Instituto de Geof\'isica, Unidad Michoac\'an, Universidad Nacional Aut\'onoma de M\'exico, Antigua Carretera a P\'atzcuaro 8701, Ex-Hacienda de San Jos\'e de la Huerta, 58190 Morelia, Michoac\'an, M\'{e}xico}
\address[5]{School of Information Technology, Deakin University, Melbourne, 3125, Australia}
\address[6]{Department of Mathematics, Physics and Electrical Engineering, Northumbria University, Newcastle upon Tyne NE1 8ST, UK}
\address[7]{Instituto de F\'{\i}sica y Matem\'{a}ticas, Universidad Michoacana de San Nicol\'as de Hidalgo. Edificio C3, Cd. Universitaria, 5840 Morelia, Michoac\'{a}n, M\'{e}xico}

\received{--}
\finalform{--}
\accepted{--}
\availableonline{--}
\communicated{S. Skirvin}

\begin{abstract}
%%%
Studies on small-scale jets' formation, propagation, evolution, and role, such as type I and II spicules, mottles, and fibrils in the lower solar atmosphere's energetic balance, have progressed tremendously thanks to the combination of detailed observations and sophisticated mathematical modelling. This review provides a survey of the current understanding of jets, their formation in the solar lower atmosphere, and their evolution from observational, numerical, and theoretical perspectives. First, we review some results to describe the jet properties, acquired numerically, analytically and through high-spatial and temporal resolution observations. Further on, we discuss the role of hydrodynamic and magnetohydrodynamic instabilities, namely Rayleigh-Taylor and Kelvin-Helmholtz instabilities, in jet evolution and their role in the energy transport through the solar atmosphere in fully and partially ionised plasmas. Finally, we discuss several mechanisms of magnetohydrodynamic wave generation, propagation, and energy transport in the context of small-scale solar jets in detail. This review identifies several gaps in the understanding of small-scale solar jets and some misalignments between the observational studies and knowledge acquired through theoretical studies and numerical modelling. It is to be expected that these gaps will be closed with the advent of high-resolution observational instruments, such as Daniel K. Inouye Solar Telescope, Solar Orbiter, Parker Solar Probe, and Solar CubeSats for Linked Imaging Spectropolarimetry, combined with further theoretical and computational developments.
%%%%
\end{abstract}


\end{frontmatter}

%% For linenumbers
%\linenumbers

 
% \newcommand\jj[1]{{\color{blue} JJ: \bf #1}} 

% Definitions for the journal names
\newcommand{\adv}{    {\it Adv. Space Res.}} 
\newcommand{\annG}{   {\it Ann. Geophys.}} 
\newcommand{\aap}{    {\it Astron. Astrophys.}}
\newcommand{\aaps}{   {\it Astron. Astrophys. Suppl.}}
\newcommand{\aapr}{   {\it Astron. Astrophys. Rev.}}
\newcommand{\araa}{    {\it Ann. Rev. Astron. Astrophys.}}
\newcommand{\ag}{     {\it Ann. Geophys.}}
\newcommand{\aj}{     {\it Astron. J.}} 
\newcommand{\apj}{    {\it Astrophys. J.}}
\newcommand{\apjl}{   {\it Astrophys. J. Lett.}}
\newcommand{\apjs}{   {\it Astrophys. J. Supp.}}
\newcommand{\apss}{   {\it Astrophys. Space Sci.}} 
\newcommand{\cjaa}{   {\it Chin. J. Astron. Astrophys.}} 
\newcommand{\gafd}{   {\it Geophys. Astrophys. Fluid Dyn.}}
\newcommand{\grl}{    {\it Geophys. Res. Lett.}}
\newcommand{\ijga}{   {\it Int. J. Geomagn. Aeron.}}
\newcommand{\jastp}{  {\it J. Atmos. Solar-Terr. Phys.}} 
\newcommand{\jgr}{    {\it J. Geophys. Res.}}
\newcommand{\mnras}{  {\it Mon. Not. Roy. Astron. Soc.}}
\newcommand{\nat}{    {\it Nature}}
\newcommand{\na}{    {\it New Astron.}}
\newcommand{\pasp}{   {\it Pub. Astron. Soc. Pac.}}
\newcommand{\pasj}{   {\it Pub. Astron. Soc. Japan}}
\newcommand{\pre}{    {\it Phys. Rev. E}}
\newcommand{\solphys}{{\it Solar Phys.}}
\newcommand{\sovast}{ {\it Soviet  Astron.}} 
\newcommand{\ssr}{    {\it Space Sci. Rev.}} 
\chardef\us=`\_

\section{Introduction}
Observations and studies of the physical properties of different types of solar jet-like structures have been carried out for more than a century. A solar jet is typically defined as collimated, beam-like features in the solar atmosphere, reminiscent of an ejection of plasma. They are essential in aiding the understanding of physical processes occurring in the solar atmosphere, such as magnetic reconnection, particle acceleration, plasma instabilities, and wave generation, to name a few. Therefore, it is crucial to understand the generation and evolution of solar jets since they may act as natural conduits for mass, momentum, and energy transport to the upper solar atmosphere. Most importantly, studies of their nature could help find answers to outstanding open questions in solar physics, particularly the physical mechanisms behind coronal heating and solar wind acceleration. Jets and jet-like eruptions are observed ubiquitously in the Sun's atmosphere, both on disk and limb. Jet activities can occur on a wide range of spatial and temperature scales in the solar atmosphere, ranging from massive eruptions in the solar corona to smaller-scale plasma jets commonly observed in the lower solar atmosphere such as the chromosphere. The chromosphere is a poorly understood layer in the solar atmosphere, mainly due to uncertainty behind the governing physical processes and the complex nature of the plasma within this region. In particular, jet activities in the lower solar atmosphere have also needed to consider physical effects related to the partially ionised character of plasmas. The temperature of the plasma in the photosphere and low chromosphere is not hot enough for the gas to be fully ionised; in reality, the plasma is a mixture of charged particles (positive ions, electrons, and neutral atoms) interacting through collisions. Therefore, ideal MHD describing the behaviour of a fully ionised plasma environment may not provide an accurate representation of the physics occurring in this region, and consequently, the effect of neutrals must be taken into consideration. This phenomenon is discussed at various points in this review. For a recently detailed overview of partially ionised plasmas in solar and broader astrophysical context, we refer the interested reader to the review by \citet{Ballester2018SSRv}.

At larger spatial-scales, jets are observed in coronal X-ray and EUV passbands and are often found near the edges of active regions (ARs) and within coronal holes \citep{shi1992}. These jets are further classified by their temperature, e.g. H$\alpha$ surges \citep[$10^{4}~\mathrm{K}$;][]{Jibben2004}, EUV jets \citep[$\sim 10^{5-6}~\mathrm{K}$;][]{Nistico2009}, and X-ray jets \citep[$\sim 10^{7}\mathrm{K}$;][]{Shibata1992, Cirtain2007}. A precedent review with a particular focus on X-ray and EUV jets can be found in e.g. \citet{Innes2016}.

Our review mainly focuses on observational and numerical findings related to smaller-scale solar jets and is structured into the following sections:  Section \ref{Observations} provides an overview of the essential observations of small-scale plasma jets. Section \ref{formation} discusses the proposed formation mechanisms for the variety of observed solar jets considered in this review. Section \ref{evolution} outlines the possible plasma instabilities that could occur during the evolution of a jet's propagation. In Section \ref{waves} we analyse the reported observations of waves and oscillations in small-scale solar jets. Where appropriate, we include discussions about results based on numerical simulations. Finally, Section \ref{summary} contains a brief discussion about the main results and suggests possible future topics of research.

\section{Properties and observations}\label{Observations}
Depending upon the location (on-disk or at limb), solar jets can be observed across a wide spectral range for time scales that vary from a few seconds to minutes. They are known as, for example, spicules, fibrils, mottles, rapid blue/red-shifted excursions (RBE/RREs), and straws. These features effectively bridge the lower solar atmosphere to the corona through the transition region, where the plasma-$\beta$ varies from very large to very small, passing through unity where the sound and Alfv\'{e}n speeds are equal. Many observations support the idea that these chromospheric jet-like features can be modeled as thin, dense magnetic flux tubes that can act as a waveguide for (transverse) oscillations and provide a natural conduit for energy and momentum transfer into higher layers of the solar atmosphere \citep{Dep2021}. There is a wide array of nomenclature for these jet-like features; however, many represent the same physical structure, depending upon where the feature is observed (e.g., on disk or at the limb) and the instrument used for the observation (Swedish Solar Telescope (SST), Richard B. Dunn Solar Telescope (DST), Hinode/SOT, etc.).

\subsection{Spicules}\label{spicules}
\begin{figure*}
	\centering
	\includegraphics[width=0.79\textwidth]{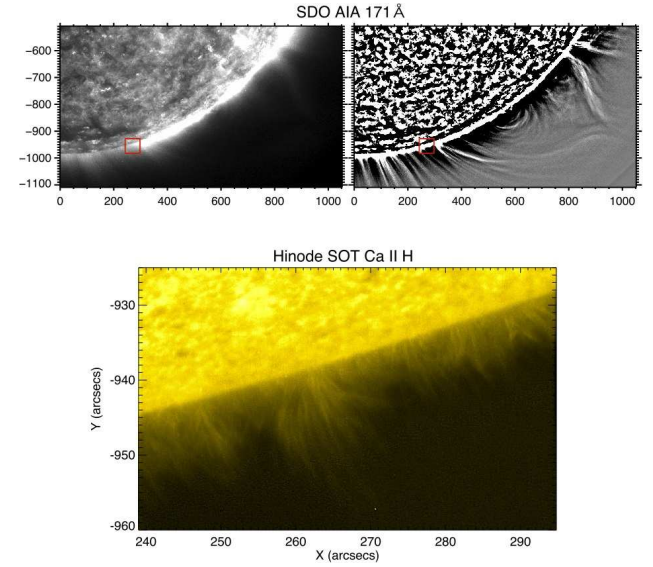}
	\caption{Image taken from Figure 1 in \citet{Okamoto_2011}. The top two panels show SDO/AIA 171{\AA} with a zoom-in region showing the field of view (FOV) of Hinode/SOT. The top right panel shows the sum of 10 consecutive images that have been treated with unsharp masking to highlight faint structures off limb. The bottom panel shows Hinode/SOT Ca \textsc{II} H image at the same time. This figure powerfully shows the ubiquity of spicular features in the lower solar atmosphere.}
	\label{fig:okamoto_spicules}
\end{figure*} 

\begin{figure*}
	\centering
	\includegraphics[width=0.79\textwidth]{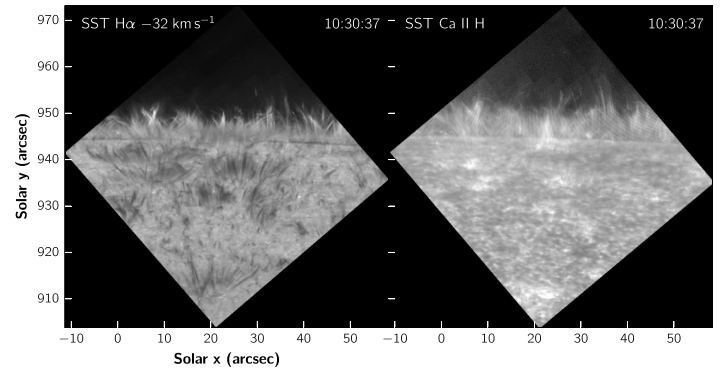}
	\caption{Image taken from Figure 1 in \citet{Pereira2016} showing the prevalence of spicules in the chromosphere. The left panel shows an image from CRISP (SST) in the blue wing of H$\alpha$. The right panel shows a simultaneous image using the Ca \textsc{II} H filter.}
	\label{fig:pereira_spicules}
\end{figure*} 

Historically, spicules \cite[see, e.g.][]{beck1968, Sterling_2000, tsi2012} have been the most routinely observed features at the solar limb and are most readily identified in chromospheric spectral lines such as H$\alpha$, Ca \textsc{II} h/k, Mg \textsc{II} and Si \textsc{IV} \citep{Pereira2016}. Spicules are most easily observed off the solar limb, as they are not optically thick enough to stand out from the photospheric background, see Figures (\ref{fig:okamoto_spicules}) and (\ref{fig:pereira_spicules}). Solar spicules appear as thin, grass-like, inclined structures that populate the lower solar atmosphere in a strong (AR) and weak e.g., quiet Sun (QS) magnetic environment and are rooted in the chromospheric network. Although spicules were first noted in the scientific literature by \citet{Secchi_1878}, understanding their physical origin has been evading researchers in solar physics ever since. 

There is much-heated debate over the classification of spicules. However, it is widely accepted that spicules can be separated into sub-classes based on their observed properties or generation mechanisms. The two sub-classes, type I and type II, may also form the basis of an interesting discussion about whether or not spicules can be classified as jets. Type I spicules are considered the `classical' spicule and have properties consistent with the fundamental physical description of jets. These type I spicules follow a parabolic trajectory, have lifetimes of roughly $5-15$ minutes and reach maximum heights of $5-10$Mm. For an in-depth summary of the properties of type I spicules, we refer the reader to the reviews by \citet{beck1968, Beckers_1972}. Type I spicules are most commonly observed in ARs, QS regions, and coronal holes. 

On the other hand, the properties of type II spicules are significantly different from their classical counterpart. In particular, magnetic reconnection driven type II spicules reach maximum heights of $3-9$ Mm (longer in coronal holes) and have shorter lifetimes of $10-270$ s as compared to their type I counterparts in Ca \textsc{II} passband (see e.g. \citet{De_Pontieu_et_al_2007a, Pereira_et_al_2012}). Observations of type II spicules also show that they undergo transverse oscillations with velocity amplitudes of order 10-20 km s$^{-1}$ with periods of $1 - 10$ minutes have been determined using Monte-Carlo simulations due to the short observed lifetimes of these structures \citep{dep2007, Tomczyk_et_al_2007, Zaqarashvili&Erdelyi_2009, McIntosh_et_al_2011, Sharma_2017}, with apparent propagation speeds of 30-110 km s$^{-1}$. These were interpreted as upward, downward, and standing Alfv\'en waves by the authors \citep{Okamoto&De_Pontieu_2011, Tavabi_et_al_2015}, however it was later shown that the waves detected were actually signatures of fast magnetoacoustic kink waves \citep{vand2008,goossens2009}. Observations by \citet{De_Pontieu_et_al_2009, De_Pontieu_et_al_2011} indicate those type II spicules are continuously accelerated as they rise through the solar atmosphere while being heated to temperatures similar to that of the transition region. Further observations suggest that some type II spicules also show an increase or a more complex velocity dependence with height in the solar atmosphere \citep{Sekse_et_al_2012}. At the end of their life, type II spicules tend to exhibit rapid fading in chromospheric lines \citep{dep2007}, further evidence of them being heated beyond chromospheric temperatures. 

The arguable non-parabolic trajectory of type II spicules and their rapid fading seen in chromospheric absorption lines have raised the debate of whether they can be classified as plasma jets. Type II spicules can be solely a radiative phenomenon of plasma moving in and out of a filter, which is used to observe the chromosphere. This phenomenon will have no vertical mass transport, but mechanical energy transport will be. Also, it explains why they can be accelerated upwards against the intense gravitational field. Their observed rapid fading out of chromospheric filters can be due to several reasons, and their temperature is just one of them. 

In addition, to type I and type II spicules, there are also large-scale counterparts, very long spicules that are also abundant in the solar atmosphere, called macrospicules. These giant features were described by \citet{Bohlin_et_al_1975}, describing the jets observed in Skylab's extreme ultraviolet (EUV) spectroheliograms at the solar limb. Besides, they are observed in polar coronal holes, where they reach heights from 7 to 70 Mm above the solar limb, maximum velocities from 10 to 150 km s$^{-1}$ and lifetimes ranging from 3 to 45 minutes. In the QS, they represent long jets of chromospheric plasma ejected to roughly 4-40 Mm coronal heights before they fall back or fade \citep{priest_2014}. These jets have been observed with vertical flow velocities over 200 km $\textrm{s}^{-1}$, and are found to be closely linked with various magnetic processes (emergence/cancellation) and particular topological features such as loops, mini-filaments, and coronal holes \cite[see, e.g.,][]{Bohlin_et_al_1975,Withbroe_et_al_1976,Dere_et_al_1989,Karovska&Habbal_1994,Parenti_et_al_2002,Bennet&Erdelyi_2015,Kiss_et_al_2017,Sterling_2000,Wilhelm_2000,Loboda&Bogachev_2019}.

\subsection{Disk counterparts of spicules}
\subsubsection{(Dynamic) fibrils}

\begin{figure*}
	\centering
	\includegraphics[width=0.79\textwidth]{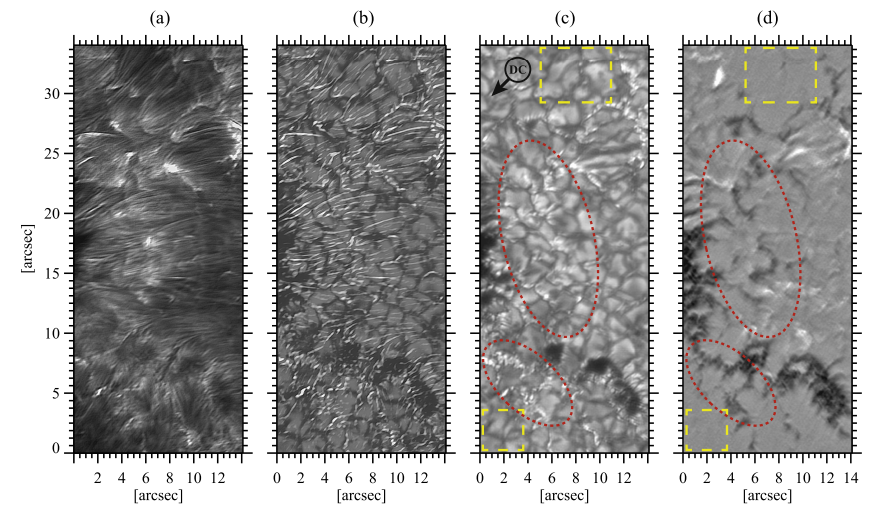}
	\caption{Image is taken from Figure 2 in \citet{Jafarzadeh2017} in which the authors investigate slender Ca \textsc{II} H fibrils using data from Sunrise balloon-borne solar observatory. Panel (a) shows a Ca \textsc{II} H image with numerous fibrils fanning out from AR located near the bottom and left of the FOV. Panel (b) shows a superposition of the identified slender Ca \textsc{II} H fibrils on top of the Ca \textsc{II} H image. Panel (c) is the same as (b); however, shown at full darkened contrast. Finally, panel (d) shows the corresponding Stokes V magnetogram, where the red ellipses outline the two regions with magnetic concentrations, whereas the yellow boxes indicate regions of QS areas.}
	\label{fig:jafarzadeh_fibrils}
\end{figure*} 

H$\alpha$ observations of the solar disk reveal that the chromosphere is replete with two types of `fibrils.' Long fibrils tend to emanate from sunspot regions and arch over super granular cells. Shorter fibrils are shorter-lived and show up as dark features in the QS seen in Figure (\ref{fig:jafarzadeh_fibrils}). These features appear dark because they reach the optical depth unity at greater altitudes than their surroundings. These features are believed to be type I spicules observed on disk and have been called `dynamic fibrils.' In QS regions, type I spicules have been interpreted as mottles when observed on the disk \citep{De_Pontieu_et_al_2007a, tsi2012}. Dynamic fibrils have been observed to have maximum velocities ranging from $10 - 30$ km s$^{-1}$ and lifetimes ranging from $2$ to $20$ minutes \citep{Hansteen_et_al_2006, Lan2008, Priya2017}.

The on-disk counterparts of type II spicules have generated even more controversy about their nature. They are referred to as `rapid blue-shifted excursions' (RBEs) when observed on disk using the SST following a curved trajectory and appear to be heated during their lifetime \citep{dep2007fibril, RouppevanderVoort2009, Sekse_et_al_2012}. First observations of suggested type II spicules on disk were reported by \citet{tian2014} using data taken from the Interface Region Imaging Spectrograph (IRIS) \citep{IRIS2014}. It was deduced by observing rapid brightening that they were travelling upward in the plane of the sky at speeds close to $300$ km s$^{-1}$. These speeds are far greater than the observed plane of sky velocities for off-limb type II spicules observed in Ca \textsc{II}. Further observational studies presented by \citet{RouppevanderVoort2015} investigated the Doppler velocities of on-disk type II spicules and found that the associated speeds were much slower than those observed by \citet{tian2014} using the plane of sky observations.

It is strongly believed that jets in the lower solar atmosphere trace out the magnetic field lines in the magnetic network. \citet{Leenaarts2015} investigated this idea by comparing numerical simulations of fibrils with the dark structures seen in H$\alpha$ observations. They concluded that their simulations showed instances where the fibrils outline single magnetic fields and bundles of magnetic field lines. It is challenging to compare H$\alpha$ observations alone to determine whether a fibril outlines a single magnetic field. An exciting result from this study was that two types of waves were found to propagate within the jet. First, compressive longitudinal waves propagating along the magnetic field, symbolic of slowing mode waves, existed and were steeped into shocks in higher layers where the plasma-$\beta$ approached unity. A second, nearly incompressible wave mode, propagating at the local Alfv\'{e}n speed was seen to propagate in all directions; however, the authors were unable to conclude if this wave mode displayed torsional wave-like behaviour.

\citet{Jafarzadeh2017} conducted a statistical analysis of slender bright fibrils in Ca \textsc{II} H pass-band using the balloon-borne observatory Sunrise \citep{Barthol2011SoPh}. The authors discuss the first reported ubiquitous transverse oscillations in slender features in the lower solar atmosphere. The authors found that these fibrils possessed properties similar to type II spicules from all data analysed. Further discussion about the nature of the observed transverse oscillations shows that the transverse displacements and velocity amplitudes are smaller than those of similar features seen in the upper chromosphere. The authors proposed that this is due to observations on disk rather than off-limb as on-disk observations are more sensitive to the lower solar atmosphere where the gas density dominates.

Recent high-cadence and high-spatial-resolution observations of dynamic fibrils using combined data from the Atacama Large Millimeter Array (ALMA) and IRIS \citep{Chintzoglou2021Apj} further suggest a parabolic trajectory of these jet-like features and also evidenced their multi-thermal nature. The authors suggested that these dynamic type II spicules (dynamic fibrils) simultaneously eject plasma at different temperatures. In the same work, numerical data from Bifrost simulation \citep{Gudiksen2011} further supported this observation. {\it Bifrost is a numerical code based on the Oslo Stagger code \citep{Nordlund&Galsgaard_1995, Galsgaard&Nordlund_1996}, which solves the standard magnetohydrodynamic (MHD) partial differential equations (PDEs) on a Cartesian grid. Additionally, it has implemented a realistic equation of state (EOS), the effect of Spitzer thermal conductivity, the radiative transfer, the effects of partial ionization through ion-neutral interactions using the generalized Ohm's law, i.e., the Hall term and the ambipolar diffusion (Pedersen dissipation) in the induction equation \citep{Martinez_Sykora_et_al_2017}}.

In addition, high-resolution images of sunspots have revealed that chromospheric super-penumbral fibrils are dominating features that span out from the umbra of a sunspot. The study by \citet{Mor2021} has shown that these super-penumbral fibrils display ubiquitous transverse motions, interpreted by the authors as the propagation of MHD kink modes. Interestingly, the motions appear to be localised to individual fibrils and not associated with the sunspot's global structure. Thus, while it is clear that (dynamic) fibrils act as an excellent waveguide for MHD waves, there remain open questions about the origin of observed motions in these features.

\subsubsection{Mottles}
\begin{figure*}
	\centering
	\includegraphics[width=0.79\textwidth]{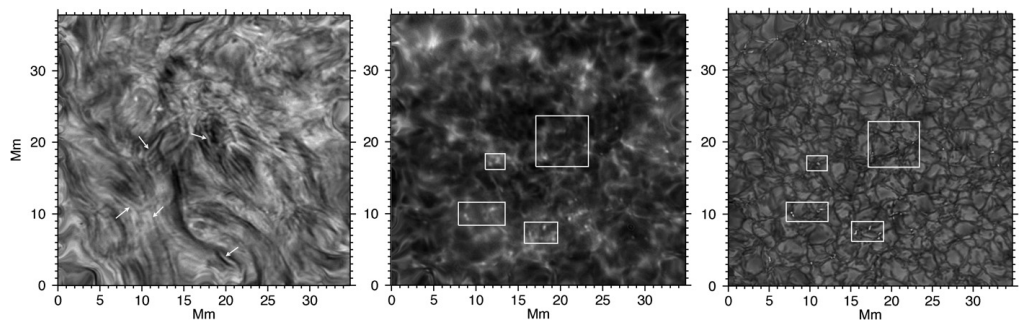}
	\caption{Image taken from Figure 1 in \citet{Kuridze_et_al_2012} showing the appearance of dark mottles on the solar disk using the ROSA (Rapid Oscillations in the Solar Atmosphere) instrument at the DST. All panels show snapshots simultaneously with the left panel imaging using H$\alpha$ line core, where the arrows show the locations of bright and dark mottles. The middle panel shows Ca\textsc{II} K core data and the right panel G band image, where the boxes indicate the locations of long-lived mottles detected with H$\alpha$.}
	\label{fig:kuridze_mottles}
\end{figure*} 

Chromospheric mottles can be classified as another form of `spicule-like features' commonly observed on the solar disk in the QS, an example shown in Figure (\ref{fig:kuridze_mottles}). The QS has features which are equivalent to AR in the sense that two types of mottles dominate the chromosphere here, namely, long horizontal dark mottles and shorter dynamic mottles \citep{dep2007mottle_inproc}. However, observations of single mottles are usually difficult because they do not possess a `top' end in observations that dominate the dynamics, unlike AR dynamic fibrils. Furthermore, mottles appear to undergo significant motions transverse to the magnetic field during their lifetime \citep{Morton2012, Kur2013}. Albeit an older study, the readers can find an excellent review of chromospheric mottles (named in the paper as disk spicules) in \citet{Sue1995}. 
It was reported in these observations that mottles were seen to follow a parabolic trajectory with both an ascending and descending phase over an average total lifetime of $2$-$15$ minutes \citep{Sue1995}. There was also a correlation between the jets' velocity and their maximum length. In the context of jet phenomena, the same study found that mottles were observed to possess velocity profiles consistent with a sudden acceleration, followed by a constant deceleration. Furthermore, a linear correlation between the lifetime of the mottle and its maximum length was revealed. These observations were also consistent with a later study by \citet{dep2007ApJmottle} in which the authors used SST data to analyse properties of dynamic fibrils and found that the two categories of chromospheric jets shared similar characteristics. However, we should note that dynamic fibrils were reported to have larger maximum lengths and lifetimes than mottles, notably due to the stronger magnetic field environment in which dynamic fibrils are located. 

\begin{table*}
%\caption{A summary of the key properties (lifetimes, maximum heights, maximum velocities, widths etc.) of spicules, mottles and fibrils from recent observational reports.}
\caption{A summary of the key properties of spicules, mottles and fibrils from recent observational reports.}
\label{Jet_properties_table}
\begin{tabular}{cccccc}     % define the column alignment
                           % l: left, c: center, r: right
  \hline                   % horizontal line
Jet type & Lifetime & Maximum height & Maximum velocity & Width & Reference \\
         & [mins] & [Mm] & [km s$^{-1}$] & [km] \\
  \hline
Type I & 3 - 10 & 2 - 9 & 10 - 30 & 700 -2500 & \cite{beck1968, Beckers_1972}\\
Type I  & 3 - 12 & 4.2 - 12.2 & 3 - 75 & 300 -1100 & \cite{Pasachoff2009} \\
Type I & 3 - 7 & 5 - 10 & -- & 120 - 700 & \cite{dep2007} \\
Type II & 0.2 - 2.5 & 1 - 7 & 50 - 150 & $\leq$ 200 & \cite{dep2007} \\
Type II & 1 - 4.5 & 5.5 - 7.75 & 30 - 70 & 300 - 350 & \cite{Pereira_et_al_2012} \\
Dynamic fibril &  0.5 - 1.5 & 4 - 10 & 80 -250 & $\leq$ 300 & \cite{tian2014} \\
Dynamic fibril &  -- & -- & 50 -75 & -- & \cite{RouppevanderVoort2015} \\
Dynamic fibril & 2 - 20 & 0.025 & ~10 - 30 & -- &\cite{Priya2017} \\
Mottle & 2 - 15 & -- & ~10 - 30 & -- & \cite{Sue1995}\\
Mottle  & 1 - 5 & 4 - 6 & 3 - 18* & -- & \cite{Kuridze_et_al_2012} \\

  \hline
\end{tabular}
\end{table*}

\section{Formation mechanisms}\label{formation}
Even though plasma jets are not yet fully understood, several possible models are proposed to explain their formation. One such model assumes magnetic reconnection as a possible mechanism; a physical process involving the interaction of two anti-parallel magnetic field lines that rearranges the magnetic topology, and magnetic energy is converted into other forms of energy, such as kinetic and thermal \citep[e.g.][]{Priest_1984, Priest_et_al_2000}. Thus, magnetic reconnection could play a key role in forming jets and contributing to the solar atmosphere's energy budget  \citep{shelyag2018_recon, Schmieder2022}. This process occurs because magnetic fields are present throughout various layers in the solar atmosphere, including beneath the visible surface, resulting in the possibility of reconnection occurring at different heights in the solar atmosphere. In particular, the chromosphere is a very dynamic environment, where magnetic reconnection can result in observable features such as H$\alpha$ upward flow events \citep{Chae_et_al_1998} and erupting mini-filaments \citep{Wang_et_al_2000}. 

One proposed driving mechanism for solar jets in the lower solar atmosphere is the so-called `whiplash effect' \citep[see, e.g.,][]{Martinez_Sykora_et_al_2017a, Martinez_Sykora_et_al_2018}. The lower layers of the solar atmosphere are (relatively) cool ($\sim10^{4}$ K) compared to the corona ($\sim10^{6}$ K); therefore, the plasma here is only partially ionized and may be modeled as a multi-fluid environment, combining the effects of charged particles and neutral atoms. The proposed `whiplash' mechanism combines magnetic tension with the effect of neutral particles found in the partially ionized lower solar atmosphere through a process known as ambipolar diffusion. The authors showed that, in their model, spicule-like features occur due to the amplification of magnetic tension and are transported upward through interaction between ions and neutrals. The tension is impulsively released to drive flows, heat plasma, and generate waves. The authors also found that the simulated type II spicules, driven by magnetic tension release, impact the corona in various manners. In particular, the complex interactions in the solar corona region can be connected with blue-shifted secondary components in coronal spectral lines (red-blue asymmetries) that were observed using the Hinode/EIS and SOHO/SUMER. Furthermore, \citet{De_Pontieu_et_al_2012} reported that type II spicules exhibit torsional motions with rotational speeds of 25-30 km s$^{-1}$. Additional numerical studies investigating the effect that a two-fluid environment include \citet{Kuzma2017a} where the authors interpret a spicule as the lifting of the chromosphere due to a signal steepening into a shock. They find that the cold, dense core of chromospheric plasma consists mainly of neutral particles and that the general characteristics of ion and neutral spicules separately are very similar to each other.

Regarding acceleration mechanisms of the formation of type II spicules, we can refer the interested reader to, for example, \citet{Martinez-Sykora_et_al_2011}, where the authors reported features found in realistic three-dimensional simulations of the outer solar atmosphere that resemble observed type II spicules. They found that the modeled spicule was composed of material rapidly ejected from the chromosphere that rises into the corona while heated. Its source lies in a region with large field gradients and intense electric currents, which lead to a strong Lorentz force that squeezes the chromospheric material, resulting in a vertical pressure gradient that propels the spicule along the magnetic field, as well as Joule heating, which heats the jet material, forcing it to fade. In addition, \citet{Goodman_2012} used a $2.5$D time-dependent MHD model to test the proposition that a Lorentz force under chromospheric conditions could generate the observed type II spicule velocities. They found that current densities localized on observed space and time scales of type II spicules and that generate maximum magnetic field strengths $\leq 50$ G can generate a Lorentz force that accelerates plasma to terminal velocities similar to those of type II spicules.
Moreover, their analysis suggested that the radial component of the Lorentz force compresses the plasma during the acceleration process by factors as significant as $\sim100$.
Additionally to the above mechanism, others are related to the siphon model. For example, \citet{Williams&Taroyan_2018} presented a mechanism associated with siphon flows that could play an essential role in prominence, filament channel, and spicule formation and their dynamics. Specifically, they found that the siphon flow undergoes a hydrodynamic (HD) shock, which allows the Alfv\'en instability to amplify the propagating waves as they interact with the shock and loop footpoints, which ultimately leads to photospheric material being “pulled” into the loop and spreading along its entirety.

Other non-ideal effects can also be responsible for the formation of chromospheric jet-like features. In the study by \citet{Yang_et_al_2013} (see their Figures 2-5), a numerical experiment was conducted including resistivity, thermal conduction, radiative losses, and an empirical heating function to simulate chromospheric anemone jets related to the moving magnetic features. The authors found that the moving magnetic features can form chromospheric anemone jets with a resulting tearing mode instability creating plasmoids that may explain the bright moving blobs in observations. Advanced MHD simulations, including non-ideal effects, have also been applied to model coronal jets. For example, \citet{Szente_et_al_2017} implemented jets into a well-established three-dimensional, two-temperature (electrons and protons) MHD solar corona model employing Alfv\'en-wave dissipation to produce a realistic solar-wind background. Mainly, the simulated jets matched with line-of-sight synthetic observations in EUV and X-ray bands. Furthermore, the key contributors to the agreement were the comprehensive thermodynamic model and the inclusion of a dense chromosphere at the base of the jet-generating regions. The two-temperature model used in \citet{Szente_et_al_2017} arises from the Alfv\'en Wave Solar Model (AWSoM) presented by \citet{van_der_Holst_2014}, which includes: (i) three different temperatures (the isotropic electron temperature and the parallel and perpendicular ion temperatures), (ii) partial reflection of the Alfv\'en waves by the Alfv\'en speed gradient and the vorticity along the field lines, (iii) the wave dissipation to the three temperatures by employing the theories of linear wave damping nonlinear stochastic heating, iv) the collisional and collisionless electron heat conduction. More recently, a numerical study by \citet{Navarro2021} modelled the generation of a jet with a Gaussian pressure driver with the inclusion of thermal conduction effects. They found that the inclusion of thermal conduction modifies many aspects of the resulting jet, including its morphology. They further found that the jet appears more collimated and can penetrate deeper into the solar corona. The inclusion of thermal conduction in their model also increased the energy and mass fluxes carried by jet. A 2D numerical study was conducted, which analysed the differences between adiabatic and non-adiabatic MHD on the evolution of spicules triggered by a velocity pulse \citep{Kuzma2017b}. The authors find that as the velocity pulse propagates into the less dense chromosphere, it steepens into a shock, with its non-linear wake causing quasi-periodic rebound shocks. The authors deduce from their model that the thermal conduction and radiative cooling terms do not significantly influence the dynamics and evolution of their generated spicular features.

\begin{figure}
\centerline{\includegraphics[width=0.49\textwidth,clip=]{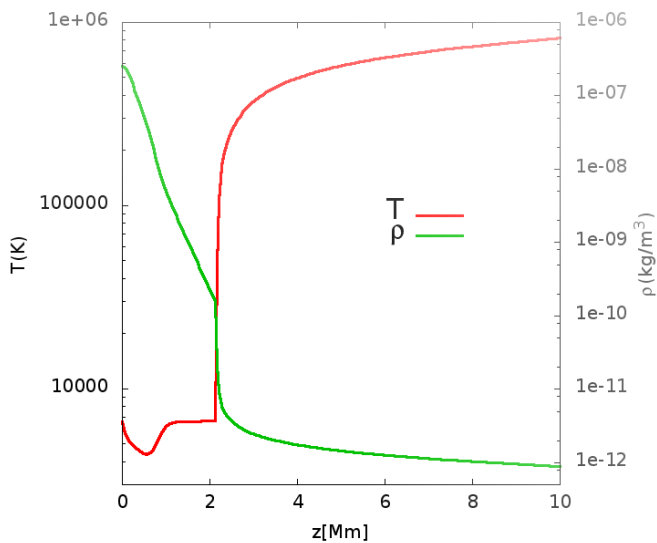}}
\caption{Equilibrium profiles for the C7 model: temperature (red line) and mass density (green line) as function of height $z$. Note the steep gradients near the transition region ($z\sim$2.1 Mm). Courtesy \citep{Gonzalez-Aviles_et_al_2017}, Figure 1.}
\label{fig:atmosphere}
\end{figure} 

\begin{figure*}
\centering
\includegraphics[width=4.8cm,height=6cm]{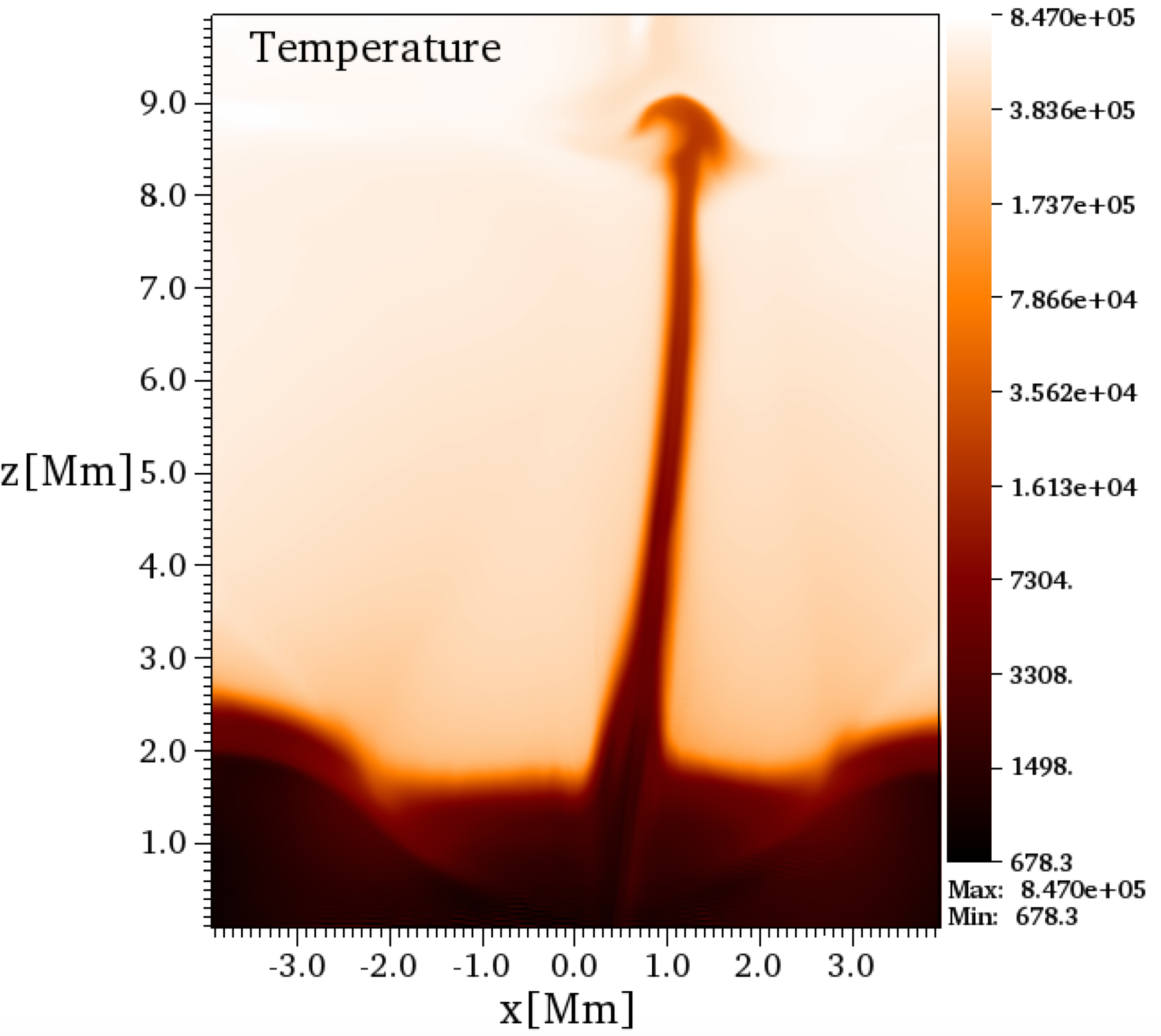}
\includegraphics[width=4.8cm,height=6cm]{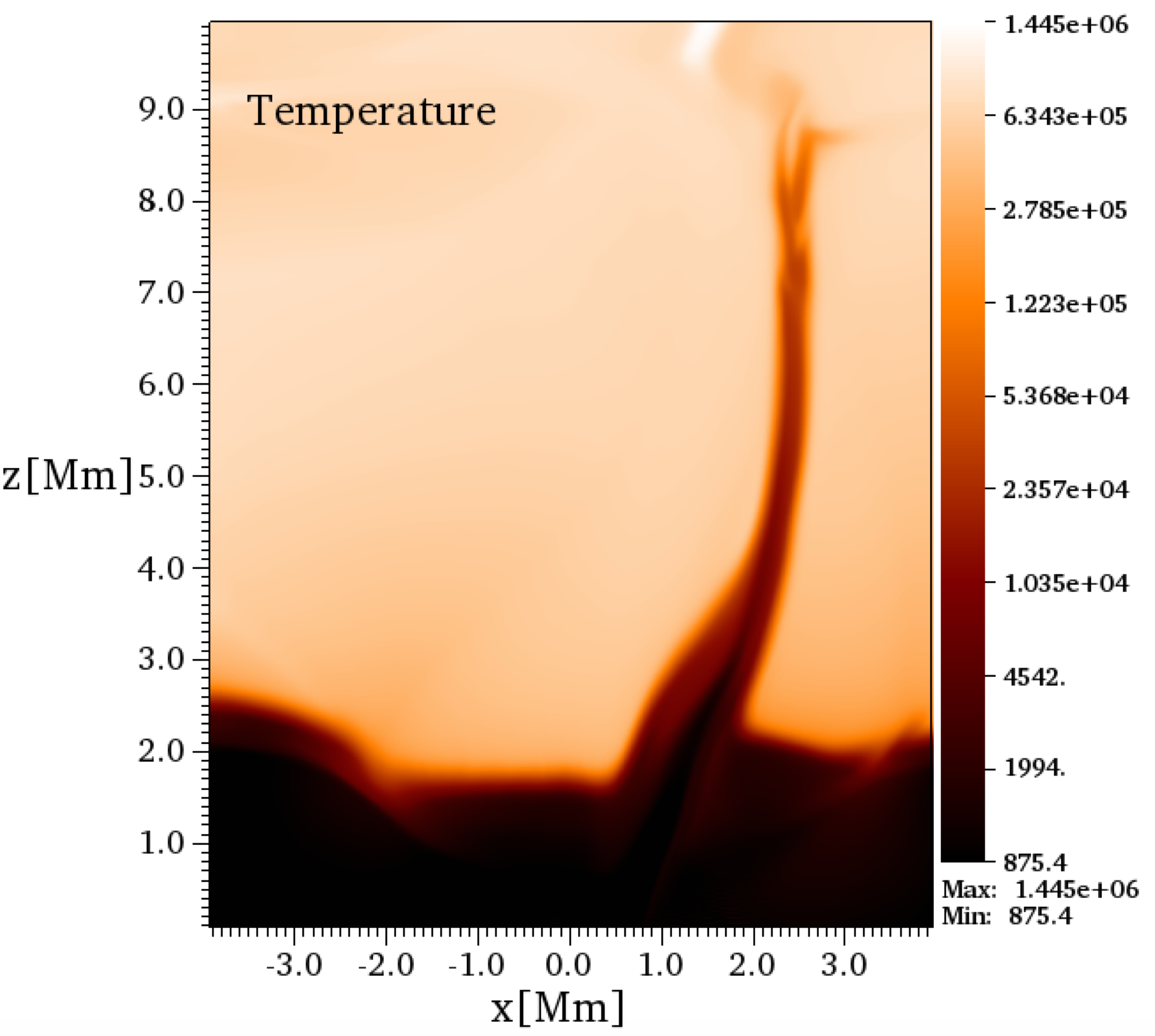}
\includegraphics[width=4.8cm,height=6cm]{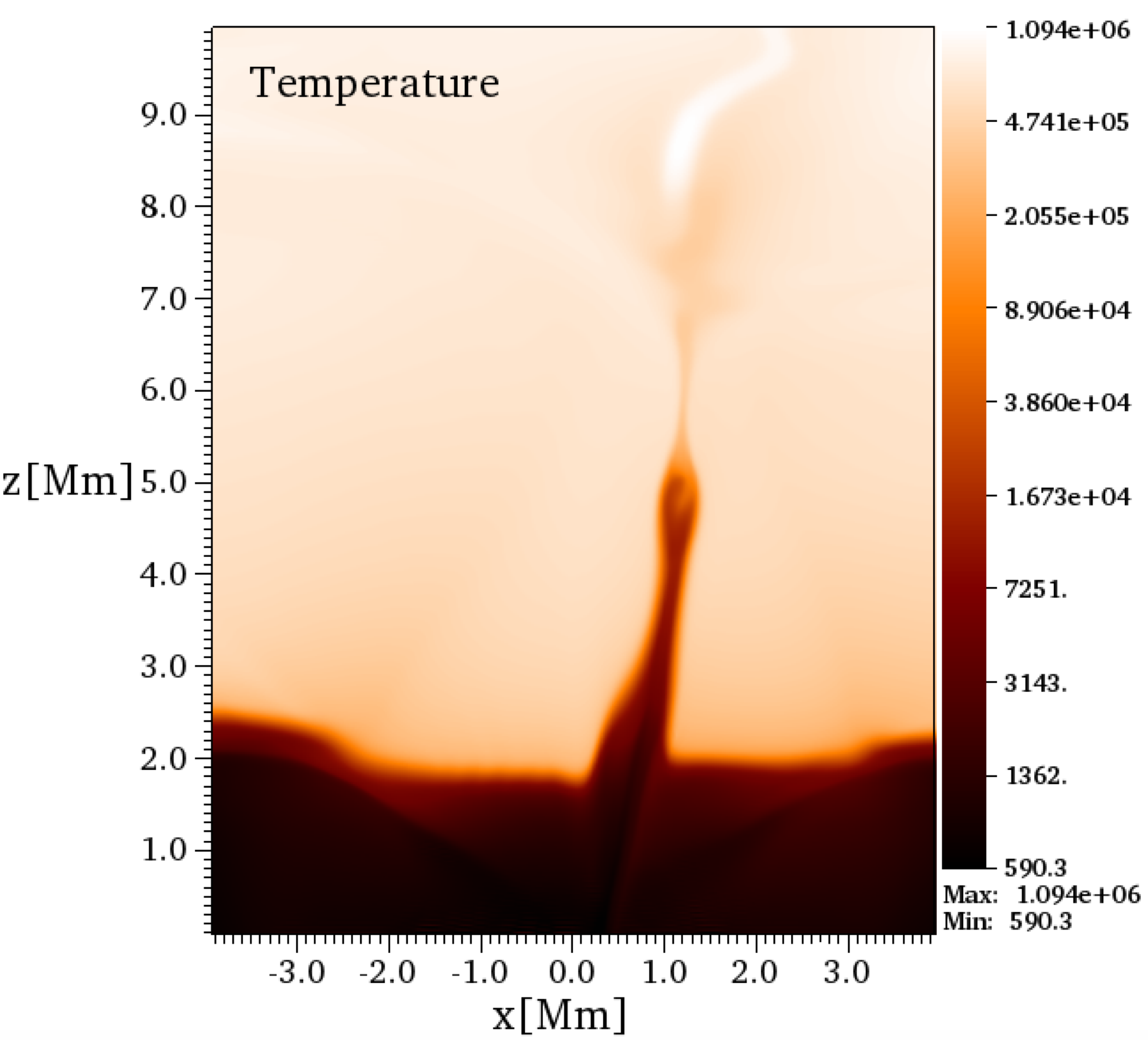}
\caption{From left to right panels show snapshots of the variation of temperature (measured in Kelvin) obtained  by numerically simulating a solar jet for three different magnetic field strength at the foot point of the left loop $B_{01}$ and the right loop $B_{02}$: (i) $B_{01}=40$, $B_{02}=30$ G at time $t=213.3$ s; (ii) $B_{01}=40$, $B_{02}=20$ G at time $t=211.2$ s; and (iii) $B_{01}=30$, $B_{02}=20$ G at time $t=204.8$ s. Courtesy \citet{Gonzalez-Aviles_et_al_2017}, Figure 6.}
\label{fig:caseB1}
\end{figure*}

A recent study by \cite{Gonzalez-Aviles_et_al_2017} proposed a model where the resistive magnetic reconnection may be responsible for the formation of jets with characteristics of type II spicules and cold coronal jets. The numerical domain used for their simulations covers the region from the photosphere to the solar corona. Figure \ref{fig:atmosphere}, shows the typical profiles of the logarithms of temperature and mass density for the standard C7 solar atmospheric model \citep{Avrett&Loeser2008} that describes the properties of the chromosphere-transition region, and it is extended to the solar corona \citep{Fontela_et_al_1990, Griffiths_et_al_1999}. Figure \ref{fig:caseB1} displays examples of jet structures from \citet{Gonzalez-Aviles_et_al_2017} illustrated with a snapshot of the plasma temperature for three different combinations of the magnetic field strength of loops. In the three cases, the jet appears at the center of the two loops. The authors assumed an asymmetric distribution of the magnetic field along the loop. In particular, the generated jets show an inclination towards the loop with smaller magnetic field strength. It was found that simulated jets reach maximum heights of $\approx$ 7.6 Mm measured from the transition region for the case when the magnetic field strength of the loops are $B_{01}=40$ G and $B_{02}=30$ G, respectively (the two values denote the strength of the magnetic field in the two footpoints). The jets reached heights slightly greater than the estimated maximum heights between 5.78-6.73 Mm observed in type II spicules in the QS and coronal holes \citep{De_Pontieu_2007, Pereira_et_al_2012}. In case of the same magnetic field strength of the loops, i.e., $B_{01}=B_{02}=40$ G, the jet reaches a maximum height of about 6.7 Mm, which is also similar to the observed maximum height of type II spicules.
In contrast, in the case corresponding to $B_{01}=30$ G, $B_{02}=20$ G, the obtained jet is smaller than in the two previous cases, reaching a maximum height of approximately 3.1 Mm, that is not consistent with the observational heights of type II spicules. Regarding the vertical velocities, the simulated jet of the first case reaches a maximum vertical velocity of about 16.8 km s$^{-1}$, which is below the estimated velocities (30-100 km s$^{-1}$) of the observed type II spicules \citep{Pereira_et_al_2012}. In the second case, the jet reaches a maximum vertical velocity of 31 km s$^{-1}$, which is near to the lower range of the velocities estimated for type II spicules. Finally, in the third case, the maximum velocity in the jet is slower ($\approx$ 5.8 km s$^{-1}$) than the two previous cases, and we can identify that its velocity is not similar to the observed spicules. The lifetime of the simulated jet obtained in the first case is about 400 s, which is greater than the typical lifetimes of the observed type II spicules, that is in the range of 50-150 s \citep{Pereira_et_al_2012, Skogsrud_et_al_2015}. The obtained lifetime of about 400 s has more similarities with the lifetime estimated (150-400 s) to the type I spicules \citep{Beckers_1972, Sue1995, Sterling_2000}. The other two simulated jets' lifetimes are in the same order ($\approx$ 400 s) to the jet of the first case; therefore, they are also not related to the lifetimes of type II spicules. The comparisons with the observed parameters of type II spicule are only qualitative since they do not calculate emission maps, which may help compare the results with observations more precisely. 
\begin{figure*}
\centering
\includegraphics[scale=0.2]{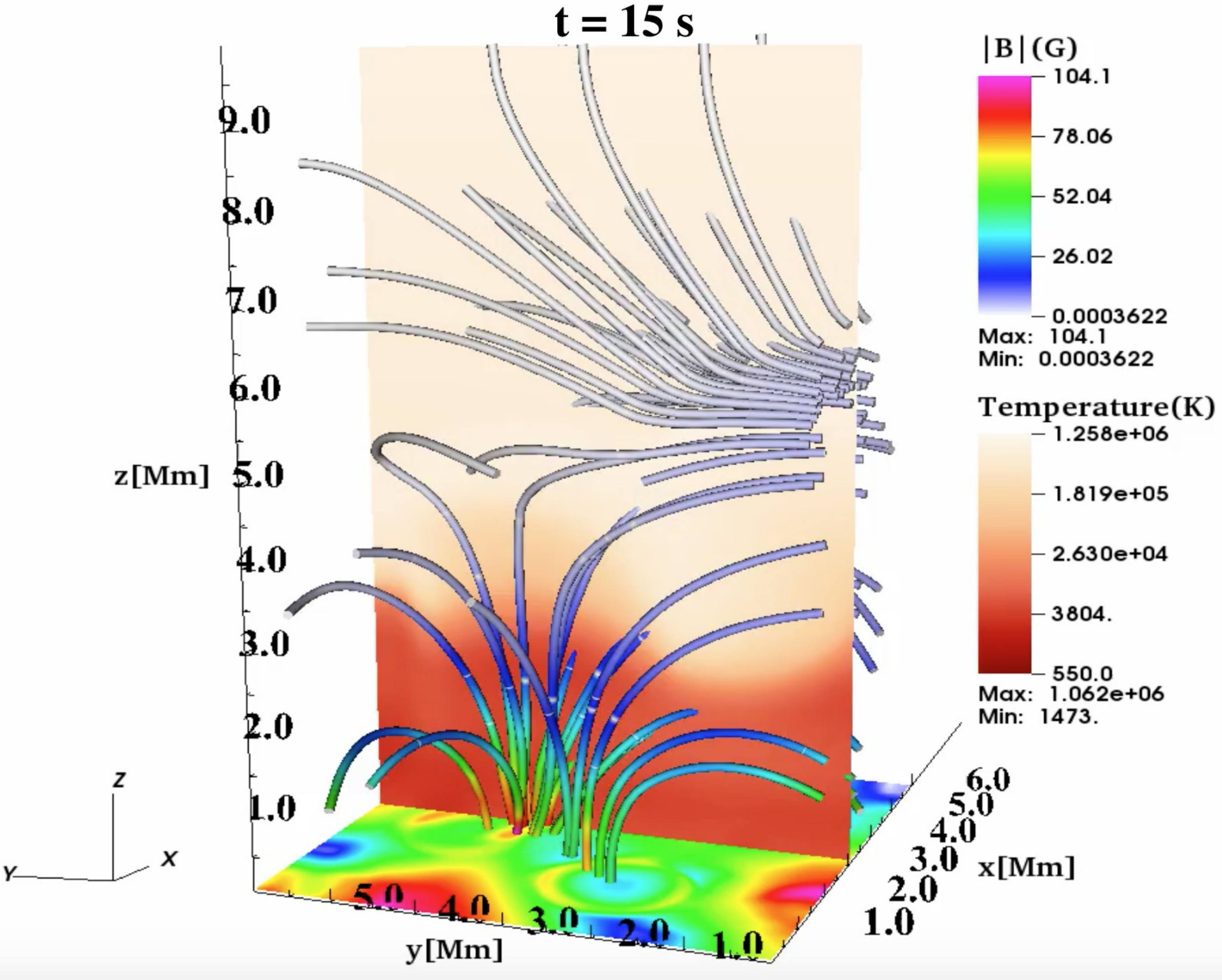}
\includegraphics[scale=0.2]{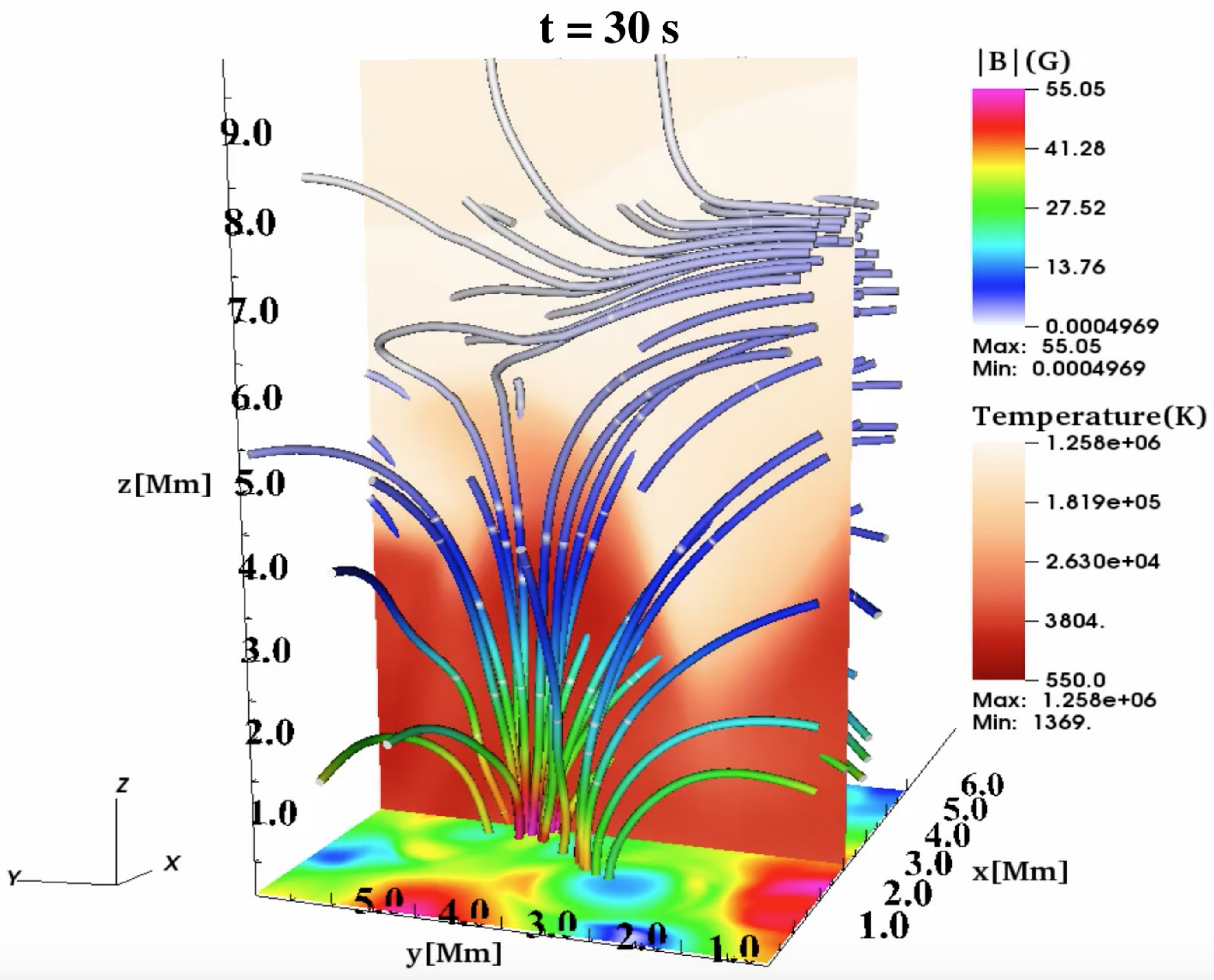}\\
\includegraphics[scale=0.2]{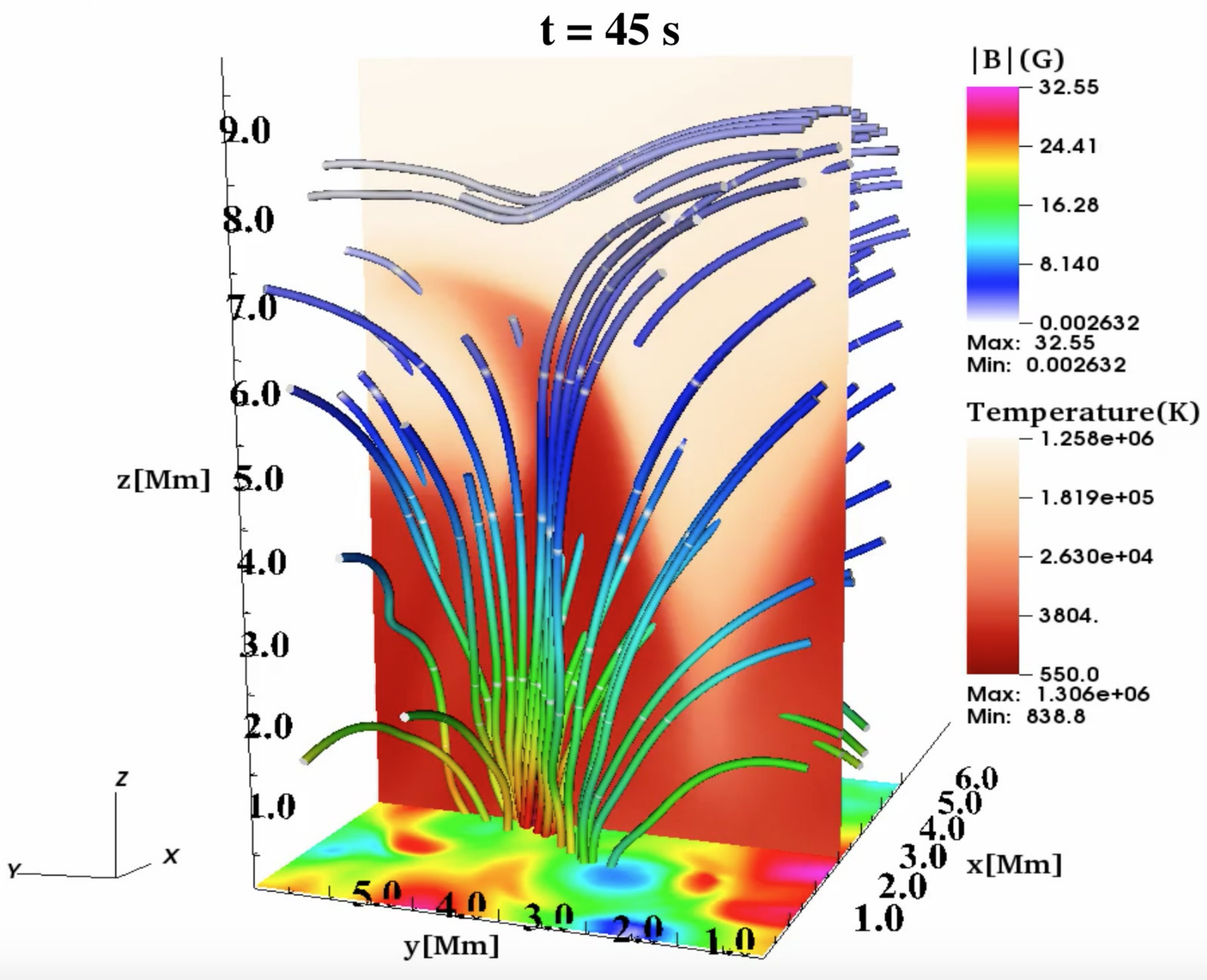}
\includegraphics[scale=0.2]{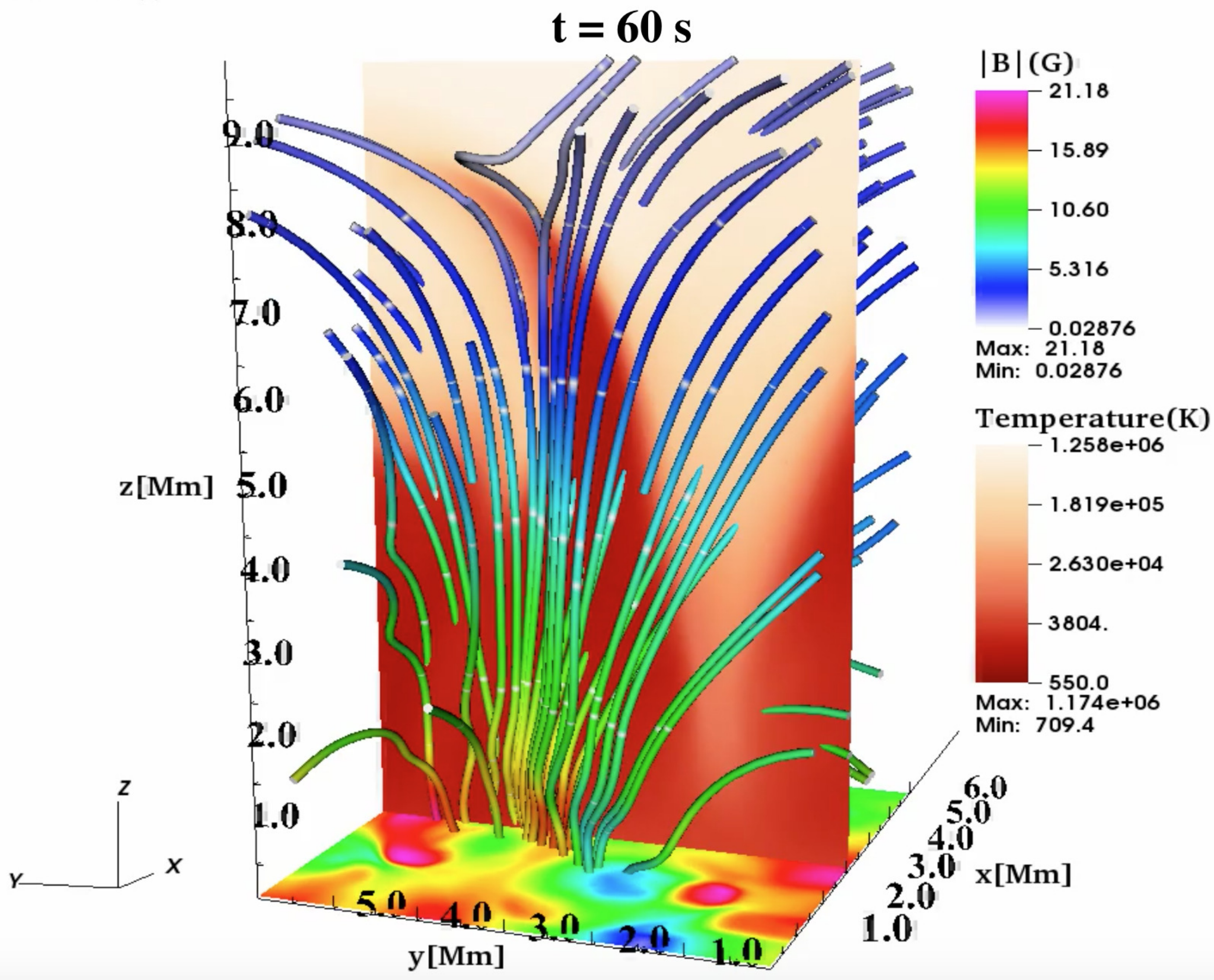}\\
\includegraphics[scale=0.2]{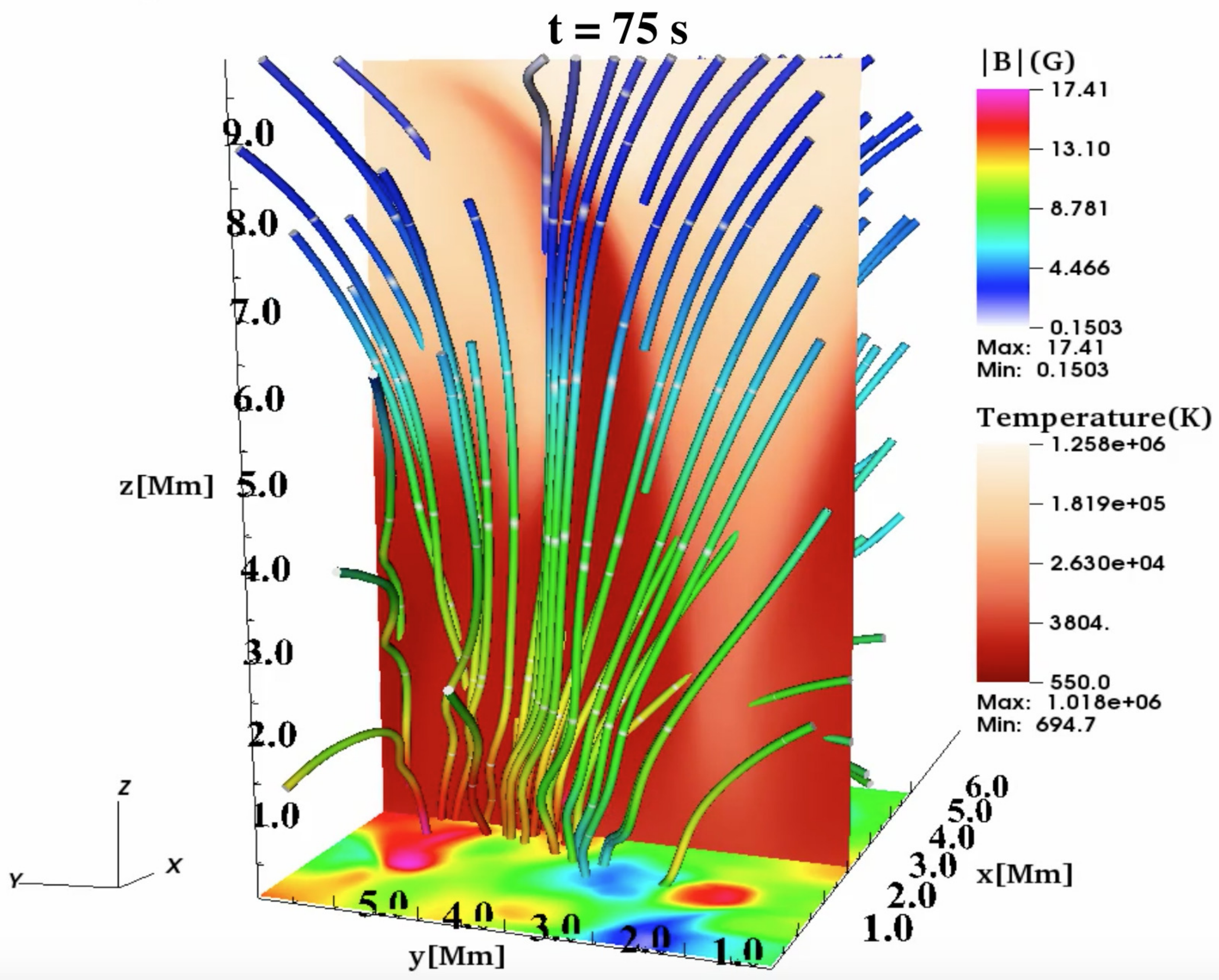}
\includegraphics[scale=0.2]{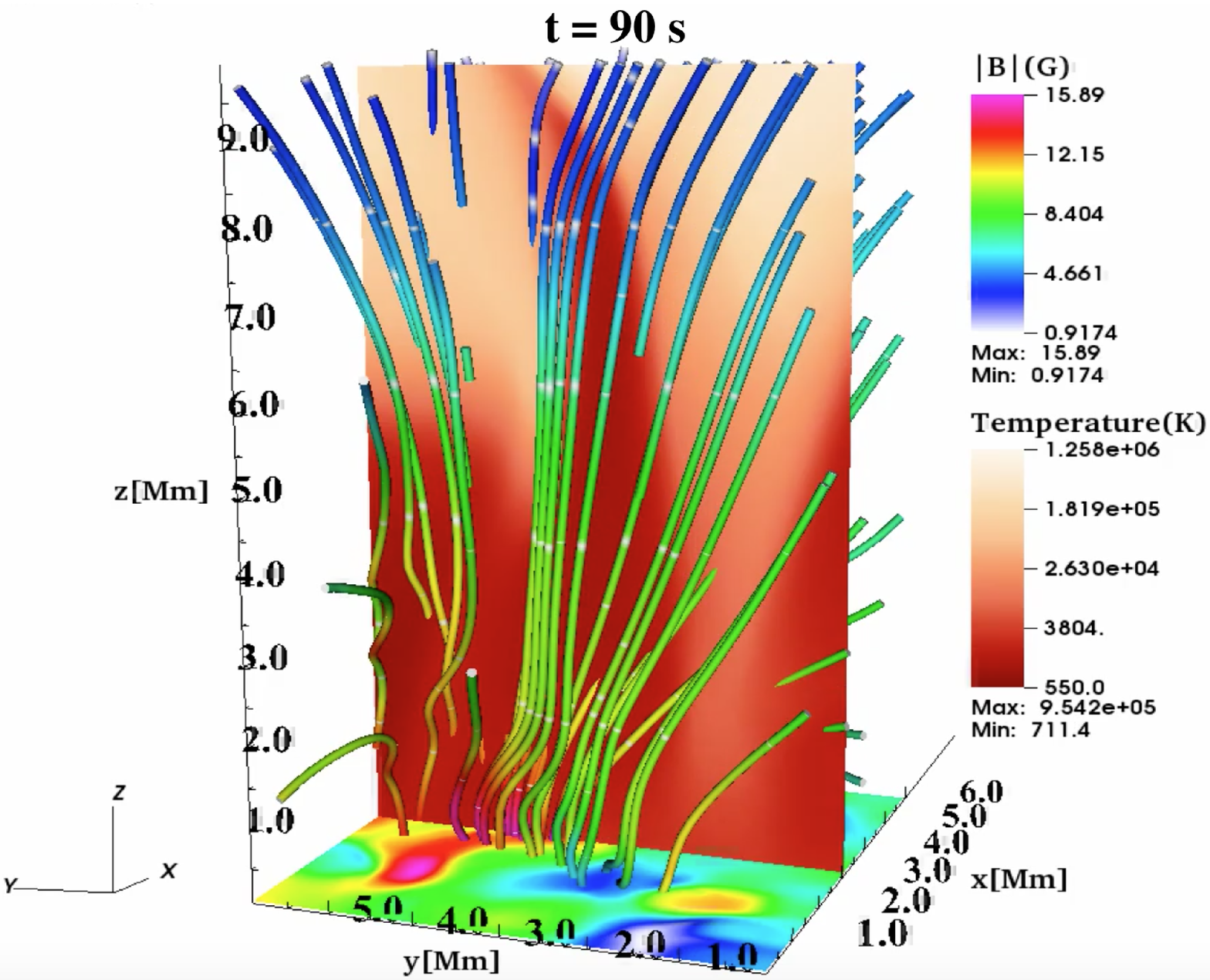}
\caption{\label{logTemp_3D_fieldlines} Snapshots of the 3D evolution of jet formation represented by a cross-cut of the temperature (K) and magnetic field lines at different times. The bottom of the domain is coloured by the magnitude of the magnetic field in Gauss. Courtesy \citet{Gonzalez-Aviles_et_al_2018}, Figure 3.}
\end{figure*}

According to the previous studies, resistive MHD is sufficient to track the reconnection process and, in general, the jet's evolution. The two-dimensional simulations by \citet{Gonzalez-Aviles_et_al_2020} have shown that the inclusion of thermal conduction along the magnetic field lines slightly affects the temperature and morphology of the spicule jets. A complementary study was to simulate a solar jet excitation in three-dimensional geometry \citep[see, e.g.][]{Gonzalez-Aviles_et_al_2018}. The authors investigated the possibility that magnetic reconnection may be responsible for forming jets similar to type II spicules.  The 3D potential magnetic field was taken from solar magnetoconvection simulation performed with the MURaM code \cite[see e.g.][]{Vogler_et_al_2005, Shelyag_et_al_2012} and the equilibrium model is same as in \cite{Gonzalez-Aviles_et_al_2017}. The implemented model has a more complex magnetic field configuration that can mimic the complex structure of loops in a region of the solar atmosphere.
The 3D representation of the jet formation and the variation of the temperature and magnetic field lines at different times are shown in Figure~\ref{logTemp_3D_fieldlines}. The jet structure forms in a region where magnetic reconnection is triggered. The jet reaches the maximum height, which is similar to the heights of type II spicules observed in the QS and coronal holes \citep{Pereira_et_al_2012}. In addition, the jet reaches a maximum vertical velocity $v_{z}\approx 130$ km s$^{-1}$, which is higher than the estimated maximum upward velocities (30-100 km s$^{-1}$) of the type II spicules \citep{Pereira_et_al_2012}. Magnetic reconnection in \citet{Gonzalez-Aviles_et_al_2018} occurs more naturally than in the 2D simulation \citep[see, e.g.][]{Gonzalez-Aviles_et_al_2017} since the 3D magnetic field shows bipolar regions that can trigger the reconnection process, and therefore excite the formation of collimated plasma structures. Another numerical study by \citet{Smirnova2016} reproduces a spicular feature resulting from a magnetic reconnection process. The authors investigate a pressure pulse that is launched from a null point located within a magnetic arcade. They find that the resulting jet exhibits a dynamically evolving two-component structure, which is commonly observed in high-resolution observations, and mimicking both type I and type II spicules \citep[][]{De_Pontieu_2007}. They show that the pressure pulse initially creates a hot and tall jet-like structure which is then followed by a cooler denser structure as a result of evacuation of the plasma from the magnetic arcade.

Apart from magnetic reconnection, a variety of other models have also been proposed to explain the formation of spicules, including granular buffeting \citep{rob1979} and p-mode leakage \citep{dep2004}. Over the last 30 years, several studies \citep[see, e.g.][]{Shibata&Suematsu_1982, Shibata_et_al_1982, Mart2009, Guerreiro2013} have used hydrodynamic numerical simulations of processes related to the spicule jet-like formation. These studies investigate the influence of initial atmospheric structures on the dynamics of such jets.
\begin{figure*}
\centerline{\includegraphics[width=0.79\textwidth,clip=]{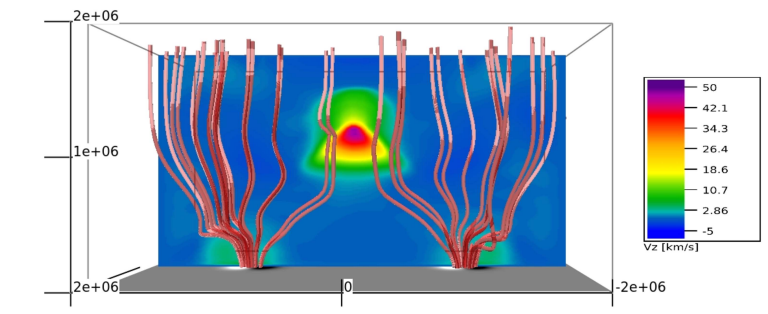}}
\caption{The vertical velocity $v_z$ color contour shows the formation of a high velocity upflows region in the center of two merging expanding flux tubes driven by counter-rotating vortex motions at the footpoints. Streamlines of the magnetic field are also shown. The resulting shock waves may drive the dynamics of chromospheric jets. Adapted from \citet{sno2018}.}
\label{fig:snow_2018_shocks}
\end{figure*} 
Recently it was shown \citep{sno2018} that shock waves might be involved in the formation of jets, generating incompressible wave modes \citep{dep2004} at greater chromospheric heights, and act as a heating mechanism of the chromosphere. The work by \citet{sno2018} considered expanding magnetic flux tubes from the photosphere into the upper solar atmosphere with counter-rotating vortex motions at the footpoints. These vortical motions stress the magnetic field of the large-scale flux tube, creating smaller structures inside. These structures behave as waveguides to the upper solar atmosphere. The interacting vortex motions create a superposition in-between the merging flux tubes, generate intense upward shocks and drive chromospheric jets (see Figure~\ref{fig:snow_2018_shocks}).  \citet{Hansteen_et_al_2006}, \citet{De_Pontieu_et_al_2007} and \citet{Heggland_et_al_2007} used numerical modeling and observations to show that dynamic fibrils, mottles, and spicules are consequences of upwardly propagating shocks passing through the upper chromosphere and transition region toward the corona. One of the most promising non-reconnection candidates is the rebound shock model, proposed first by \citet{holl1982}, who considered the possibility that acoustic-gravity waves may be able to generate shocks in the chromosphere. This process would then result in spicules manifesting as the upward motion of the chromosphere and transition region.
In their model, an initial pulse excites acoustic-gravity waves in the photosphere and propagates along magnetic flux tubes. Due to the rapid decrease in chromospheric/coronal plasma density, these waves eventually steepen into shocks, thrusting the transition region upwards, with the resulting upward-moving plasma interpreted as a spicule. The scenario of a rebound shock considers that the initial wave pulse propagates upwards and leaves behind a wake once the shock has formed (in the upstream region). Consecutive shocks are formed in the stratified atmosphere due to this wake and can continue to lift chromospheric plasma, which is observed as spicule jets \citep{zaq2011AIP}. The horizontal widths of the resulting spicules are interpreted from the presence of chromospheric plasma in the corona. The amplitude of the initial pulse directly affects the interval between the rebound shocks. Smaller amplitudes result in $\sim$3 min periods, the acoustic cut-off period in the solar atmosphere.
Similarly, larger initial pulses will result in longer periods. One consequence of this model is that it may also account for the observed multi-threaded nature of spicules. In this model, cold chromospheric plasma is lifted into the hot corona, falling under gravity and interacting with more rising plasma due to the rebound shocks. This interaction between rising and falling plasma portions may create the multi-threaded nature of spicules.

Numerical simulations of the rebound shock model have accurately reproduced many of the observed characteristics of type I spicules and macrospicules under adiabatic and non-adiabatic conditions \cite[see, e.g.][]{zaqmur2010, Mur2011, Gonzalez-Aviles_et_al_2021}. In these studies, the authors modelled the dynamics of spicules by solving the 2D MHD equations. These simulations reasonably well reproduce the observed speeds, widths, heights, and periodicities of type I spicules, as well as their multi-threaded nature. However, numerical investigations into the rebound shock model do not reproduce the characteristics of type II spicules, suggesting that separate physical mechanisms are responsible for the formation of type II spicules. Most theoretical models used to describe the formation of type I spicules are adaptations of the rebound shock model, which consider the same processes of wave amplification and shock-transition region interactions. 

Another jet formation mechanism was suggested by \citet{holletal1982, dep1998, kud1999, jam2002}. In their so-called Alfv\'{e}n wave model, the Alfv\'{e}n waves may be non-linearly coupled to fast magnetoacoustic shocks. This model suggests that the nonlinear Alfv\'{e}n wave is converted into the acoustic wave, which is compressible and, as a result, can steepen into shocks. As the torsional Alfv\'{e}n wave amplitude is zero at the axis of the waveguide, the accelerated parallel plasma flow must have an annulus or "macaroni"-like structure in the perpendicular direction \citep{Shestov2017}. A 3D radiative MHD simulation has been presented by \cite{iij2017}, which models the formation of spicule-like jets by the Alfv\'{e}n waves. Their numerical modelling utilises the photosphere's detailed radiative transfer, which produces thermal convection that can ultimately excite MHD waves. The resulting waves create jets with similar characteristics to type I spicules. The simulations presented by these authors have the advantage that the driving mechanism, the observed multi-stranded structure of spicules, and their oscillatory nature are all explained in the same model. 

\begin{figure*}
\centerline{\includegraphics[width=0.79\textwidth,clip=]{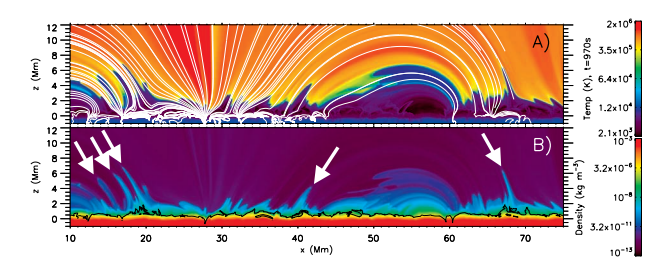}}
\caption{Taken from \citet{Martinez_Sykora_et_al_2017a}. The resulting spicular features (indicated by arrows) were due to radiative MHD simulations in the lower solar atmosphere. The spicule-like features show up as cool and dense intrusions into the solar corona. Panel (a) shows the temperature contour, and panel (b) shows the density contour. The magnetic field is over-plotted with white contours indicating regions where the plasma-$\beta$ equals $1$.}
\label{fig:mart_sykora_2017_spicules}
\end{figure*} 

Radiative MHD simulations by \citet{Martinez_Sykora_et_al_2017a, Martinez_Sykora_et_al_2018} reproduce some features resembling type II spicules (see Figure~\ref{fig:mart_sykora_2017_spicules}). However, these MHD simulations struggle so far to accurately reproduce the observed characteristics of type I spicules, such as heights, lifetimes, speeds, densities, and temperatures. One aspect not covered in these studies is the dissipation of MHD waves due to the small spatial scales required to investigate these phenomena. These numerical studies provide initial evidence that the physical discrepancies observed between the type I and type II spicules are due to different driving mechanisms. As noted above, the rebound shock model's consecutive shocks provide a reasonable estimate for type I spicules; however, the method based on the release of magnetic tension more accurately reproduces type II spicules' observed characteristics. Recent 3D numerical simulations of chromospheric jets \citep{Quintero2019MNRAS} have synthesised the spectral line profiles from MHD magnetoconvection simulations to be compared with observations. The authors investigated the synthetic Stokes profiles of the obtained jet and found that the twist of the jet impacts the magnetic field, causing the field lines to arch and bend around the core of the jet.

Although significant large-scale reconnection events such as solar flares and coronal mass ejections (CMEs) have a different morphology than smaller-scale features such as X-ray and EUV jets, they do share similar characteristics \cite[see, e.g.][]{wyp2017}. Theoretically, these events were considered to arise from different physical mechanisms; however, observations have suggested that coronal jets may be generated by the same processes which drive CMEs \cite[see, e.g.][]{ste2015}. Therefore, studying and understanding large-scale reconnection events could help provide valuable insight into the smallest scale features' dynamics at the limit of currently available spatial resolution. While reconnection is an essential aspect of the formation of solar jets, an in-depth review of magnetic reconnection processes is not the aim of the current paper. For more details on the reconnection process, the authors would like to suggest the reviews by \cite{Priest2007}, \cite{YamRevModPhys}, and \cite{Wie2014}.

\section{Jet evolution}\label{evolution}
The previous section has detailed how the jets formed; however, we have not yet pointed out how they evolved. Therefore, this section describes jet evolution and the dynamical phenomena associated with it. In particular, we can observe oscillations due to perturbations to the initial equilibrium. These perturbations may grow in time in the form of various instabilities. For example, MHD waves in the solar atmosphere may become unstable, and the nature of the resulting instability depends upon the particular type of wave mode. Also, the kink mode may become unstable due to the Kelvin-Helmholtz instability (KHI). Furthermore, the high propagation velocities in which some solar jets are characterised by maybe a natural catalyst for the onset of KHI.

\subsection{Kelvin-Helmholtz instability}
The KHI is a fundamental instability commonly found in various aspects of hydrodynamics (HD) and MHD \citep{cha1961, ofm2011, zhe2015rev}. This instability arises due to a velocity shear across a perturbed boundary, separating two fluids. The KHI is characterised by vortices appearing at the sheared interface, where the plasma mixes through turbulent motion. Spectral line broadening can infer the resultant turbulent plasma motion and non-thermal heating. Unlike the purely HD scenario, in MHD, a magnetic field aligned with plasma flow acts to stabilize the flow \citep{cha1961}, suppressing the onset of the KHI. However, it is also known that a magnetic field perpendicular to the flow does not hinder KHI development. 
The KHI has been investigated as a possible source of heating in coronal loops due to phase mixed Alfv\'{e}n waves \citep{Hey1983}. The phase mixing theory is based on structures with a perpendicular gradient in the local Alfv\'{e}n speed resulting in strong velocity gradients with Alfv\'{e}n waves propagating independently along separate field lines. In a coronal loop structure, where standing shear Alfv\'{e}n waves are highly likely to be found, they are subject to KHI, resulting in strong velocity gradients due to multiple reflections at the loop footpoints. Thus, the KHI can help sheared Alfv\'{e}n waves to reach a state of turbulence, forming small length scales in the velocity and magnetic fields, greatly enhancing wave dissipation and potentially contributing to coronal heating loops \citep{brow1984}. Furthermore, the study by \cite{Sri2021} showed that KHI could act as a possible contributor to the energy balance of the solar chromosphere. 

Further to the KHI, solar plasma jets are also subject to kink instability (due to the transverse motion of the jet axis) which depends on the Alfv\'{e}n Mach number (the ratio of Alfv\'{e}n speed to flow speed) \citep[see, e.g.][]{zaq2020}. For a sufficiently large jet velocity, the jet can become unstable if the resulting centripetal force overcomes the Lorentz force of the magnetic field. The authors further showed that the kink instability might be more pronounced for spicules that rise to the upper regions of expanding flux tubes. At these heights, the spicules become inclined and maybe more unstable with higher velocities. Finally, the authors hypothesised that KHI might drive the kink instability and, therefore, this may explain the rapid fading of type II spicules in chromospheric spectral lines.

It is well known that a straight magnetic field parallel to the interface dividing two separate fluids suppresses the onset of KHI. However, \citet{bar2019} proposed that a magnetic flux tube with even a small magnetic twist parameter is always unstable to KHI. The authors showed that magnetic shear could not stabilise an interface between oppositely directed flows. Nevertheless, it may reduce the instability growth rate under certain conditions \citep[see][for more details]{bar2019}. The results showed that magnetic shear could explain the development of KHI at the boundary of oscillating magnetic flux tubes by taking into account the azimuthal component of the magnetic field. 

 \citet{zaq2010} analysed the importance of the presence of an axial mass flow in a magnetically twisted tube for reduction of the threshold of the kink instabilities. The authors have shown that sub-Alfv\'{e}nic flows of solar jets might be sufficient for the development of KHI, provided there is enough magnetic twist. \cite{zaq2014} investigated the effect of external magnetic field on the development of the KHI, and they found that an axial external magnetic field will suppress KHI onset, whereas even a slight twist in the external field will allow KHI development for sub-Alfv\'{e}nic flows.

Numerical simulations of solar jets have provided greater insight into the possible unstable evolution of jets, which observations cannot offer at their current resolution. The 2D simulated solar jet discussed in Section \ref{formation} by \citet{Gonzalez-Aviles_et_al_2017} is an example of this. The appearance of a bulb-like structure (shown in Figure \ref{fig:caseB1}), possibly related to KHI \citep{Kuridze_et_al_2016}, is caused by the magnetic field suppressing it \citep[see, e.g.,][]{Gonzalez-Aviles_et_al_2017, Gonzalez-Aviles_et_al_2020}.

Some numerical studies were conducted to investigate the role of viscosity and resistivity on the generation and evolution of the KHI. \citet{How2017} have shown that due to the formation of KHI in a twisted magnetic field, large currents are produced, resulting in greater Ohmic dissipation and possibly strengthening the wave contribution to heating of the outer solar atmosphere. Further work by \citet{Ant2018} used an idealised 3D MHD simulation in which they modelled a spicule as a density enhanced magnetic flux tube. The authors analysed the effect transverse MHD waves have on the boundary of spicules and any associated instabilities in the magnetic structures. They found that transverse wave-induced Kelvin Helmholtz (TWIKH) rolls appear, extracting energy from the flux tube's resonant boundary layer and dissipating it through KH vortices at the edges of the spicule. \cite{Ant2018} also concluded that the TWIKH rolls could lead to the formation of multi-threaded structures of the flux tube, similar to the observed chromospheric spicules. Thus, the effect of magnetic field properties on KHI's development in solar structures is evident. However, more work is required to fully understand the development and properties of TWIKH rolls, including the properties of flows required for their generation and the specific plasma configurations needed for such structures.

The existence of a neutral component in a partially ionised plasma may invalidate the classical criterion for the onset of the KHI and may allow its development for sub-Alfv\'enic field-aligned flows. Studies by \cite{Watson2004}; \cite{Soler2012}; \cite{Martinez2015} showed that partially ionized plasmas can develop the KHI even for sub-Alfv\'enic shear flows. These authors concluded that one reason behind the lower instability threshold is the different behaviour of neutral species that is not influenced by the stabilizing effect of the magnetic field. However, the magnetic field effect appears indirectly through the collisions of neutrals with ions. Although the ion-neutral collisions reduce the instability's growth rate, it is impossible to completely suppress the onset of the KHI originated in the neutral component. These theoretical results explain the origin of, e.g., the numerical evidence of turbulent flows in various simulated chromospheric structures (jets, prominence fibrils, etc) in terms of the sub-Alfv\'enic KHI in partially ionized plasmas. 

Another instability with the same roots as KHI is the dissipative instability that arises at the interface between two media and is related to the phenomenon of negative energy waves. This instability appears for flows whose speeds are below the KH
threshold. Usually, the interface between two media allows the propagation of two modes travelling in opposite directions; however, for flow speeds larger than a critical value, the propagation direction of the two waves becomes identical. Then, the wave having the smaller phase speed becomes a negative energy wave, which means that dissipation produces an amplification of the wave amplitude, which leads to instability, while at the same time wave energy decreases. In this case, the dissipative mechanisms working in the two regions amplify this negative energy mode leading to dissipative instability. Using an interface between a partially ionised and viscous fully ionised media \cite{Ballaietal2015} showed that in the limit of weak damping, while the forward propagating wave is always stable, for the backward propagating wave there is a threshold of the flow (below the KH threshold), for which the wave becomes unstable. Their analysis also showed that partial ionisation stabilizes the interface for any degree of ionisation.

Observational signatures of the KHI have been reported in coronal mass ejections \cite[see e.g.][]{fou2011,ofm2011,zhe2015KHIcorona}), blowout jets \citep{li2018}, EUV jets \citep{bog2018}, coronal hole jets \citep{zhe2018chjet}, chromospheric jets \citep{Kuridze_et_al_2016} and surges \citep{zhe2015KHIsurge}. In all cases, the formation of vortex-like or sawtooth structures can be seen at the boundary separating regions with different flow velocities. Likewise, the plasma temperature where the KHI occurs can increase by detectable amounts. 
 
\subsection{Rayleigh-Taylor instability}
Another fundamental instability in HD and MHD is the Rayleigh-Taylor instability (RTI). It develops at the interface separating two fluids (or plasmas) of different densities when they are accelerated against each other under the influence of e.g. gravity or inertia \citep{Tay1950, cha1961}. As a result, the denser plasma is accelerated in the direction of gravitational force, and the less dense plasma moves in the opposite direction. The RTI and KHI are often closely related to each other. Acceleration of a heavier plasma against a lighter plasma under gravity can cause a velocity shear between the two, thereby exciting secondary KHI. This may result in plasma turbulence, localised heating due to plasma mixing, and plasma wave excitation. The RTI can often be identified by `finger-like' structures developing in a plasma. This instability has been observed and analysed in many astrophysical systems, such as galaxy clusters \cite[see, e.g.][]{Rob2004ApJ, Oneill2009ApJ} and supernova remnants \citep{Jun1995, Jun1996ApJ}. The RTI has been observed less frequently than the KHI within the solar context. Numerical studies \citep[see, e.g.][]{Iso2005} have suggested that the RTI may be responsible for the formation of coronal loops during flux emergence from the lower to the upper solar atmosphere. The authors conducted 3D MHD numerical simulations of an emerging magnetic flux and its consequent reconnection with the overlying coronal magnetic field. As the emerging flux carries heavier plasma higher into the atmosphere, it is susceptible to RTI development. The resulting RTI generates thin current sheets in the emerging flux where plasma is rapidly heated and accelerated.
Further numerical investigations extended this study to include effects due to ion-neutral interactions of the partially ionized lower solar atmosphere. \citet{Arb2007} investigated the effects of partial ionisation on emerging flux through the solar atmosphere and compared it against a scenario in which the plasma was taken to be fully ionised. They discovered that when the plasma is taken to be fully ionised, the emerging flux creates a unphysically low temperature in the corona. This low temperature causes the chromospheric material to rise, resulting in the onset of RTI, as discussed by \citet{Iso2005}. However, the addition of neutrals removed this low temperature region in the overlying corona, causing less chromospheric material to be lifted and hence suppressing the instability. The plasma in the lower solar atmosphere is partially ionised, and, as a result, the combined effect of ions and neutrals may change the typical KHI and RTI. The RTI in partially ionised plasmas has been studied both analytically \citep{Diaz2014} and numerically \citep{Khom2014} under an assumption that the magnetic field both in the dense prominence thread and in the surrounding hot plasma has the same direction. This restriction has been later relaxed by \citet{Ruderman2018}. Based on observations, the angle between the longitudinal axis of a prominence and magnetic field vector is of the order of $53^{\circ}\pm 15^{\circ}$ \citep{Bommier1994}. The authors assumed a magnetic shear between the interface of a heavy partially ionised prominence and a lighter, fully ionised coronal plasma. Their analysis showed that the RTI appears only to wavenumbers smaller than a critical value determined by the degree of shear and density but significantly independent of the degree of plasma ionisation. Their results also showed that the RTI is sensitive to the degree of plasma ionisation only for small values of plasma-$\beta$ and in a very weakly ionised state. For an excellent recent review of how the theory of standard instabilities has been developed for partially ionised plasmas, we refer the reader to the instability chapter of \citet{Sri2021} to avoid repeating the same studies here.

The theory of RTI has been significantly developed in laboratory plasma physics. While this paper intends to focus mainly on aspects related to the field of solar physics, here we will briefly discuss some of these findings in laboratory plasma experiments. RTI was observed routinely in laboratory experiments with plasma jets \citep{moser2012}. The analytical theory of RTI has been greatly advanced in experimental plasma physics, and there is potential in application to solar physics. Experiments by \citet{moser2012} showed that RTI developed only on one side of a magnetic cylindrical plasma jet in the presence of lateral gravity. This observation was later explained in great detail by \citet{zhai2016}, where the authors developed a new theory, which included the effects of lateral gravity on RTI. It was found that under these conditions, the RTI couples to a different current-driven instability when both gravity and the magnetic field are important. It forms a new hybrid instability that can not be described by the theory of the two instabilities alone. It has been shown that the magnetic RTI can cause magnetic reconnection, which may have implications in the theory of solar eruptions \citep{RTI2020}. The authors conducted a laboratory experiment in which a plasma jet undergoes the kink instability coupled to the RTI, and the latter is seen as a secondary instability. In this scenario, the RTI can generate a localised electric field, inside which electrons may be accelerated to very high energies.
Furthermore, recent experiments have attempted to model the RTI in a laboratory simulated coronal loop experiment \citep{zhang2020}. The authors found that increasing the strength of the axial magnetic field leads to an increase in the instability's wavelength. The authors also suggested that this result may explain the differences observed in plume dynamics in solar prominences. Variations in the magnetic field strength inside a solar prominence may have a meaningful impact on the size of the plumes observed. 

The RTI does not have extended observational evidence under solar conditions. However, solar observations of RTI mainly concentrated on prominences (see review by \citet{hil2018}) or eruptive events, such as CMEs, which launch dense material upwards that eventually falls under gravity, triggering the onset of RTI. Signatures of the RTI in such events have been reported in CMEs \citep{Inn2012} and prominence bubbles \citep{Ber2011}. Smaller-scale jets such as spicules and possibly X-Ray/EUV jets may be beyond current observational capabilities to identify any RTI present in these features. The current spatial resolution of both space and ground-based instruments is perhaps not enough to observe KH vortices in great detail either. However, with the advent of the DKIST telescope with its unprecedentedly high resolution, we will be able to observe and study waves and instabilities in much greater detail \citep{Rast2021SoPh}.

\section{Wave phenomena in solar jets}\label{waves}
MHD waves are one of the possible mechanisms to supply energy into the solar atmosphere and may contribute to chromospheric and coronal heating through damping their energy in processes that involve short lengths scales, such as resonant absorption, phase mixing, or even turbulence \citep{ErdelyiBallai2007, Klimchuk2015RSPTA, Doorsselaere2020SSRv}. Furthermore, since waves contain information about the medium in which they propagate, they can also be used as a diagnostic tool to infer values of physical quantities that cannot be directly or indirectly measured (magnitude and sub-resolution structure of the magnetic field, transport coefficients, heating/cooling functions, plasma ionisation degree, etc.). Due to the similarity with the seismology of Earth, this important branch of solar physics has been labelled as a solar magneto-seismology framework.

The observational and theoretical study of oscillations and waves in plasma jets is a vibrant research topic in solar physics. Observations taken in two different spectral lines, which are formed at two different heights, can provide information about the characteristics of the wave propagation in plasma jets \citep[see, e.g.][]{2012ApJ...746..183J}. Furthermore, calculating the phase difference between the two observations provides information about wave propagation speeds within jet structures \citep[for previous in-depth reviews focusing on waves and oscillations in jets see, e.g.][]{Zaqarashvili&Erdelyi_2009,mat2013,jess2015}. Before discussing the observations of different MHD waves in solar jets, it may be helpful to summarise their properties from a theoretical point of view.

\begin{figure*}
	\centering
	\includegraphics[width=0.85\textwidth]{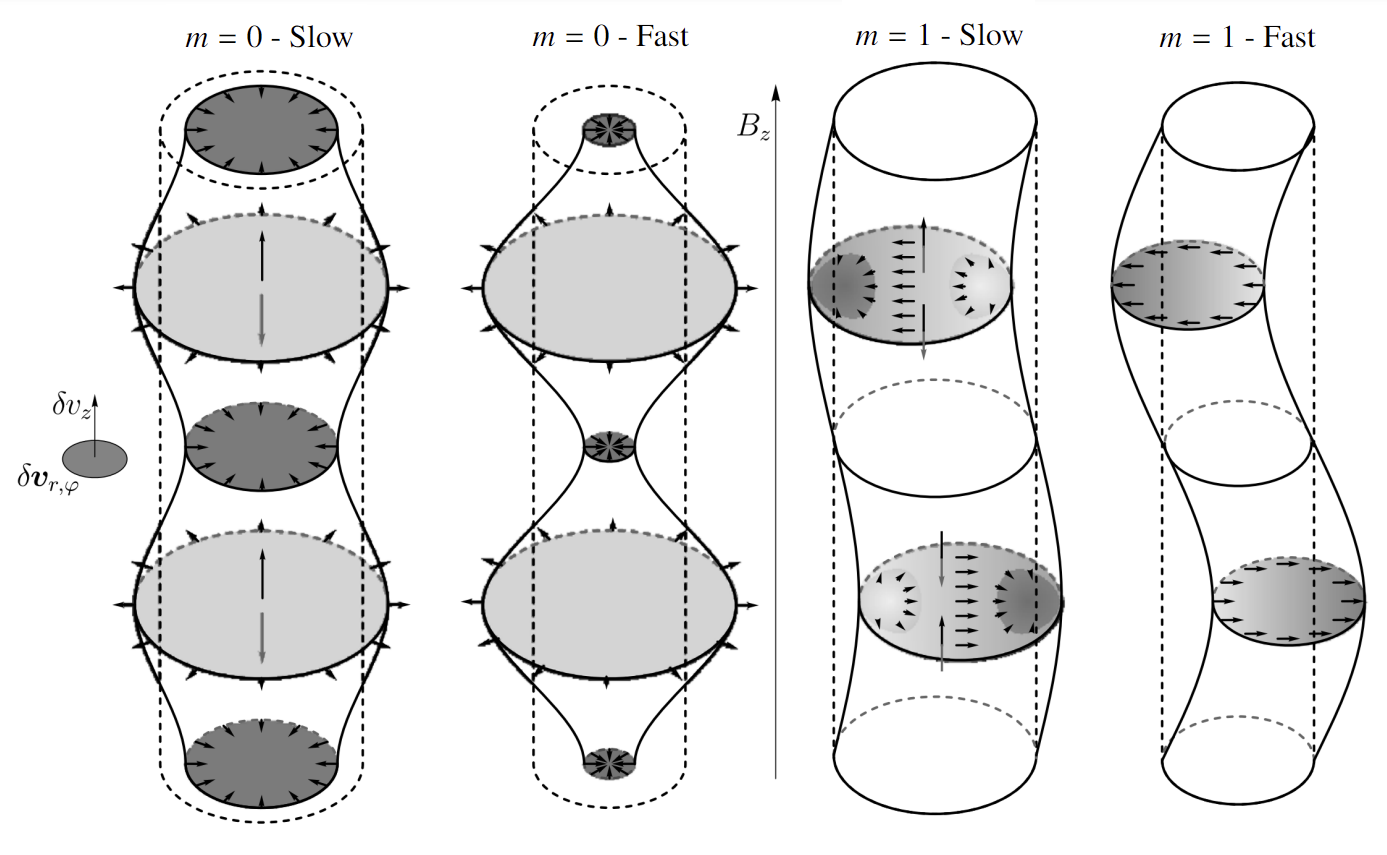}
	\caption{Schematic diagram of the magnetoacoustic wave modes observed in a magnetic cylinder \citep[adapted from][]{jess2015}. The $m=0$ sausage and $m=1$ kink mode are shown as a periodic expansion/contraction and swaying motion of the waveguide. The solid lines outline the perturbation of the cylindrical waveguide, with the corresponding arrows displaying the resulting velocity field. The magnetic field in all cases is considered vertical and uniform. A similar cartoon can be found in \citet{Morton2012}.}
	\label{fig:jess_MFT_modes_v1}
\end{figure*}

Cylindrical geometry is appropriate for the analytical modelling of MHD wave propagation in axially symmetric magnetic structures. Since waves propagate in a geometrically well-defined structure, which inherently provides an additional characteristic length-scale, they become dispersive. Oscillatory modes can be categorised in terms of their symmetry. If they propagate such that the symmetry axis of the waveguide is not disturbed, then waves are sausage waves, while if the symmetry axis undergoes a swinging motion, then we deal with kink waves. Similarly, waves with maximum amplitude on the edges of the waveguides and evanescent in the internal and external regions are labelled as surface waves. In the opposite case, if the maximum of the waves' amplitude is attained inside the waveguide, then we have body waves. A schematic diagram, which shows the magnetoacoustic wave modes in a magnetic cylinder, is shown in Figure (\ref{fig:jess_MFT_modes_v1}). In this Figure, the sausage mode corresponds to waves whose azimuthal mode number is zero, $m = 0$, inferring zero nodes in the azimuthal direction. This wave can be observed as a periodic expansion and contraction of the flux tube. On the other hand, the kink mode that corresponds to the azimuthal wavenumber $m = 1$ can be observed as a structure swaying from side to side. Likewise, in cylindrical geometry, the pure Alfv\'{e}n wave takes the form of torsional oscillations and can be observed as a periodic rotation or twisting of the magnetic structure.

\begin{figure*}
	\centering
	\includegraphics[width=0.79\textwidth]{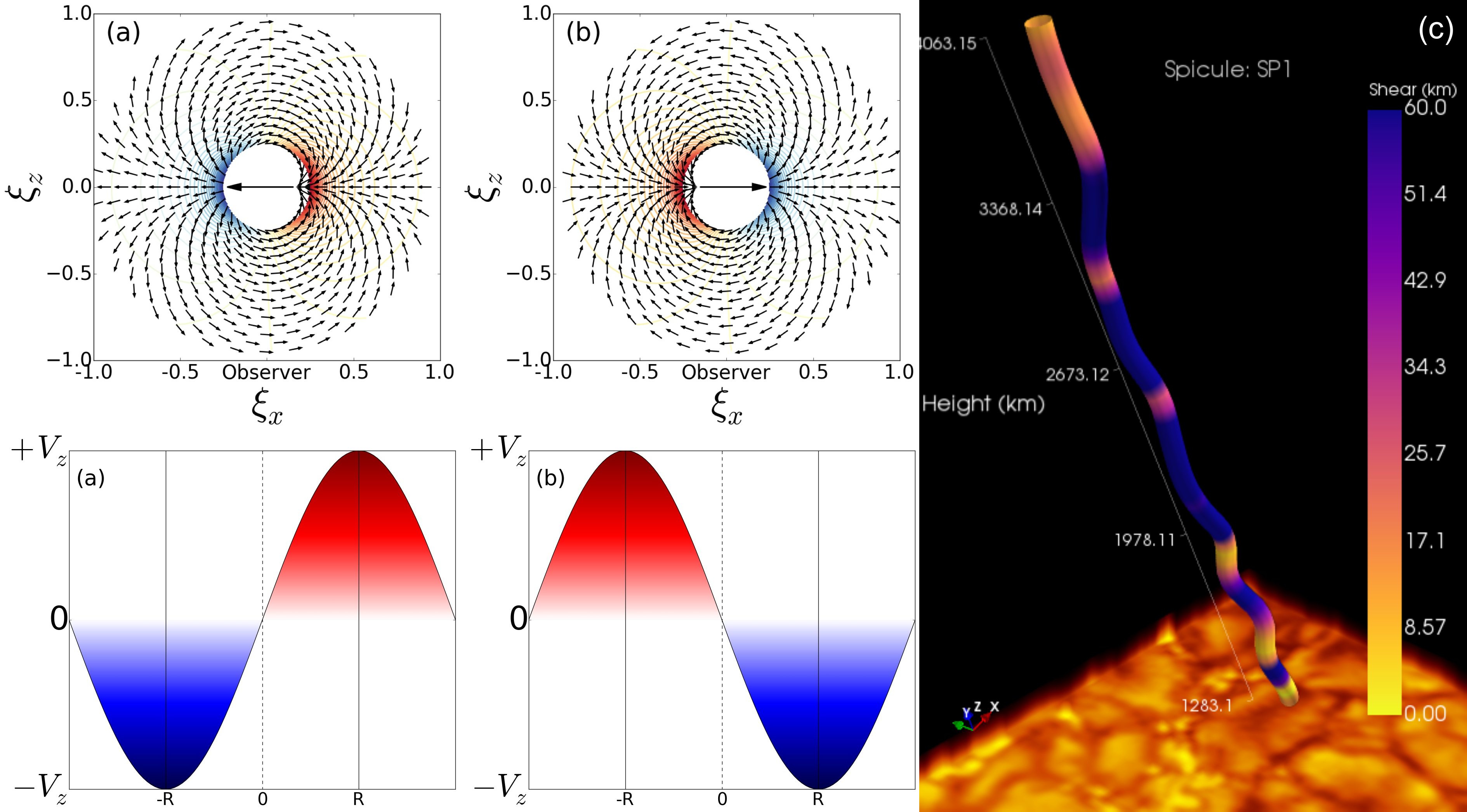}
	\caption{Adapted from \citet{Sharma2018}. Panels (a) and (b) depict a cartoon that show what Doppler profile one should expect across a spicule structure undergoing transverse displacements. (a) and (b) further show external plasma displacement field (top) of a cylindrical flux tube with radius `R' in $xz$-plane, perpendicular to the observer's line-of-sight (LOS). The waveguide's central arrow indicates the direction of the motion, due to kink ($m=1$) wave mode, with perturbed pressure color-coded around its boundary. Corresponding profiles of the red-blue axisymmetry Doppler velocities ($V_{z}$), resembling those from torsional Alfv\'en ($m=0$) wave mode, are given in the panels below \cite{Sharma_2017}. Panel (c) shows the evolution of coupled torsion and transverse motions along the height of the observed spicule feature. For (c) the width of the spicule is constant, as it is for visualization purpose only, but discussions about the variations in cross-sectional widths can be found in \cite{Sharma2018}.}
	\label{fig:velprof}
\end{figure*} 

However, since the plasma in the lower solar atmosphere is highly inhomogeneous, with waveguides having irregular geometries, MHD waves most likely have a hybrid nature, i.e., wave modes with different azimuthal wavenumber will be coupled and display mixed properties. For example, the torsional Alfv\'{e}n wave in a cylindrical flux tube can only be described as a `pure' wave when it is not coupled to another wave (e.g., kink mode), and only then the nature of the wave remains the same as it propagates along the tube. Recently \citet{Giag_2015} and \citet{Giag_2016} showed that, even in the linear regime, the presence of a magnetic background twist could couple the $m=0$ torsional Alfv\'en wave to the sausage mode, resulting in wave modes of mixed properties. 

Furthermore, due to similar spectroscopic profiles (red-blue axisymmetry), the accurate identification of observed wave modes (e.g., kink and torsional Alfv\'en) can be difficult \citep{Goossens2014, Sharma_2017}. According to \cite{Sharma_2017}, when a dense flux tube moves in the chromospheric environment, this will create perturbations in the ambient chromospheric magnetic and velocity fields. These perturbations should be, in general, the displacements of both the magnetic and velocity fields, which, if observed, can provide clues to distinguish between the dynamics of, e.g., a kink mode ($m=1$) or a torsional mode. A `pure' torsional mode will not cause a bulk transverse motion of a magnetic flux tube. The observed dynamics would be the intrinsic motion of the tube itself, with a minimal effect on the external plasma. However, if kink modes propagate along the flux tube, they will perturb the external plasma and magnetic environment, such that the external plasma would be displaced, and the surrounding magnetic field would experience compression and rarefaction. The displaced ambient plasma (in the case of a kink mode in a spicule with a circular cross-section) would have a distinctive contribution to the observed Doppler velocity. This velocity will depend on the location of the observer (or Line-of-Sight) relative to the flux tube's bulk transverse motion (see Figure~\ref{fig:velprof}) and the radiative properties of the plasma along the line of sight.  

Doppler shift oscillations have recently been observed in H$\alpha$ solar spicules \citep{Zaqarashvili2007, De_Pontieu_et_al_2012, Khut2014, Khut2017}. The authors determined the properties of spicules and calculated the oscillations present in these structures. The study by \citet{Khut2017} also revealed that the oscillations of Doppler velocity and width appear to be out of phase with one another. This anti-phase relation is determined from the strong Doppler velocity to weaker Doppler width and vice-versa. One proposed explanation for this phenomenon is due to the parabolic trajectory of type I spicules. Spicules outline the magnetic field and, as such, can be inclined to the limb and follow a parabolic trajectory, resulting in a changing velocity component along the LOS, which can cause the variation in Doppler velocity. The variation in Doppler width could be due to flow instabilities (e.g., KHI), which may lead to non-thermal broadening. Another possible explanation is that the helical motion can cause Doppler width and velocity variations due to the spicules moving along the LOS combined with the spicules' internal azimuthal motion resulting in line broadening.

Kink waves have been observed in spicules \citep{Kukhianidze2006, dep2007, Ebadi2014, Tavabi_et_al_2015}, in their on-disk counterparts (RBE/RREs) \citep{RouppevanderVoort2009}, in mottles \citep{Kuridze_et_al_2012} and in fibrils \citep{Pietarila2011}. A mix of both propagating and standing transverse waves along spicule structures was reported by \citet{Okamoto&De_Pontieu_2011}, who concluded that it is unlikely that these waves would be able to reach, and ultimately heat, the corona in sufficient quantity due to reflection at the transition region where the gradient in Alfv\'{e}n speed is high. The observational study by \citet{Sekse2013} revealed that transverse displacements along with rotational motions and mass plasma flows, a common characteristic of solar spicules, are present in RBE/RREs \citep{Kuridze_et_al_2015}. 

Several sources have been proposed to explain the existence of Alfv\'{e}n waves in the solar atmosphere, such as vortex motion in the photosphere, generated by granular and convective motions \citep{fed2011,shel2011,shelyag2013,shelyag2015}, mode conversion due to the damping of kink waves by resonant absorption \citep{goo1992, Verth2010, ter2010, goo2011} or magnetic reconnection \citep{Isobe_et_al_2008, Nishizuka_et_al_2008, He_et_al_2009, Kuridze_et_al_2012, McLaughlin_et_al_2012}. Torsional Alfv\'{e}n waves have been reported in type II spicules \citep{De_Pontieu_et_al_2012,dep2014}, inferred through Doppler images. The detected torsional motions suggest that spicules play a major role in transporting mass, energy, and helicity through the solar atmosphere by acting as waveguides for propagating waves. Another implication of this result is that the energy flux previously calculated through transverse swaying motions in spicules may be much smaller than the actual amount, as such spicules may play an even more important role in atmospheric heating than previously thought \citep{Shelyag_et_al_2012}. A detailed comparison between the transverse and rotational components of the velocity field at different heights could provide a great understanding of the dynamics, mode coupling, conversion, and damping of MHD waves in solar jets. The possibility of rotational motion in spicules to explain observed line widths was initially suggested by, e.g., \citet{Beckers_1972}. \citet{Suematsu_et_al_2008} proposed that the source of observed rotation in multi-threaded spicules is unresolved magnetic reconnection. 

Sausage waves have also been recently reported in the literature by, e.g. \citet{Gafeira2017}, who observed cross-sectional fluctuations and intensity oscillations in chromospheric fibrils on the solar disk. 

The studies mentioned above also reported wave modes occurring concurrently with one another. \citet{Jess2012} observed the kink and sausage modes in spicules and interpreted them as the transverse oscillatory and cross-sectional width variations corresponding to the kink and sausage modes, respectively. The disk counterparts of spicules were also reported by \citet{Morton2012} to exhibit the same characteristics by \citet{Jess2012}. The authors suggested that the waves are excited independently at the magnetic flux tube footpoints. These modes can then undergo mode-coupling and experience mode-conversion once they reach the plasma-$\beta=1$ layer in the transition region. \citet{Sharma_2017} suggested that depending upon the observer's LOS concerning the transverse motion of the spicule structures, they could appear to rotate either axisymmetrically or non-axisymmetrically. Furthermore, \citet{Sharma2018} identified coupling between transverse, cross-sectional width, and torsional motion observed in spicules that could be due to nonlinear processes.

Coronal seismology is used to infer plasma and field parameters of the solar corona based on the properties of observed coronal waves and oscillations. Similarly, spicule seismology \citep{verth2011} is a technique employed to estimate the variation of plasma density, magnetic field strength, and ionization degree of hydrogen along spicules that cannot be easily measured directly. The study by \cite{zaq2007} analysed $H{\alpha}$ data of off-limb spicules  using discrete Fourier  transforms and measured the spicule oscillations from Doppler velocities. Using the wave information, the authors estimated a magnetic field strength around  12-15 G at the height of 6 Mm. These authors also provided observational evidence for the 3-minute oscillations in spicules. 

\begin{figure*}
\centering
\includegraphics[width=5.5cm]{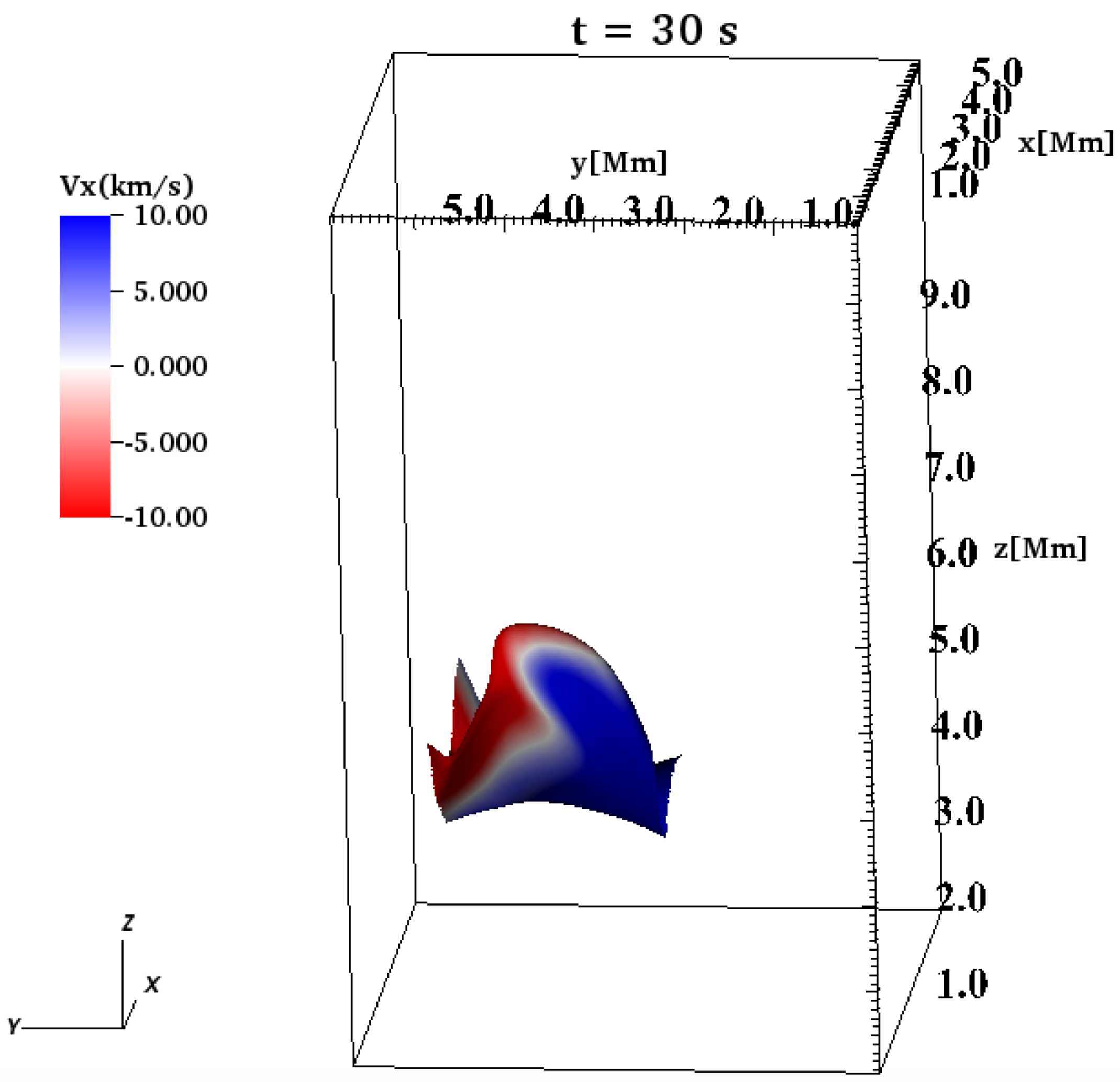}
\includegraphics[width=5.5cm]{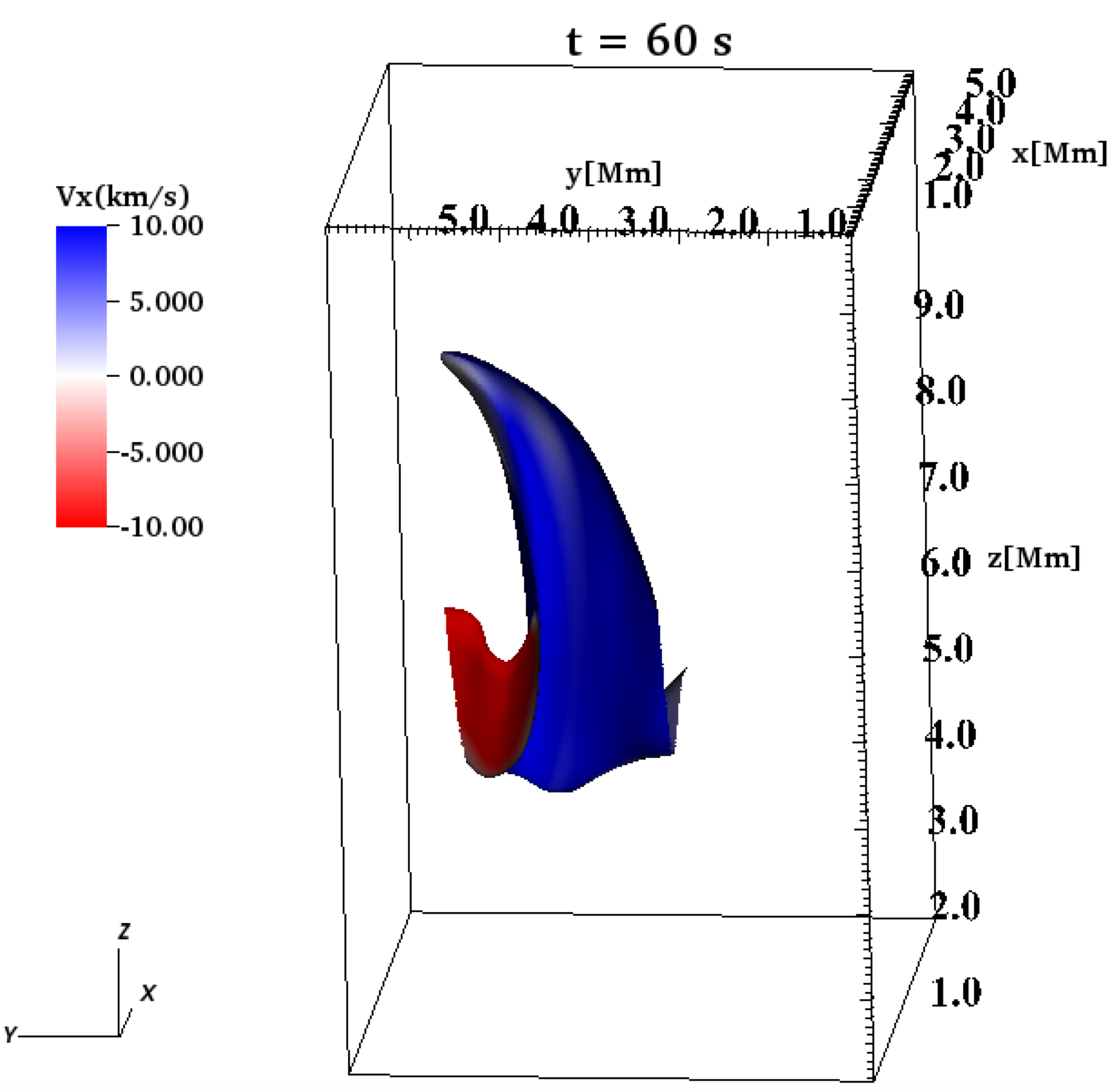}\\
\includegraphics[width=5.5cm]{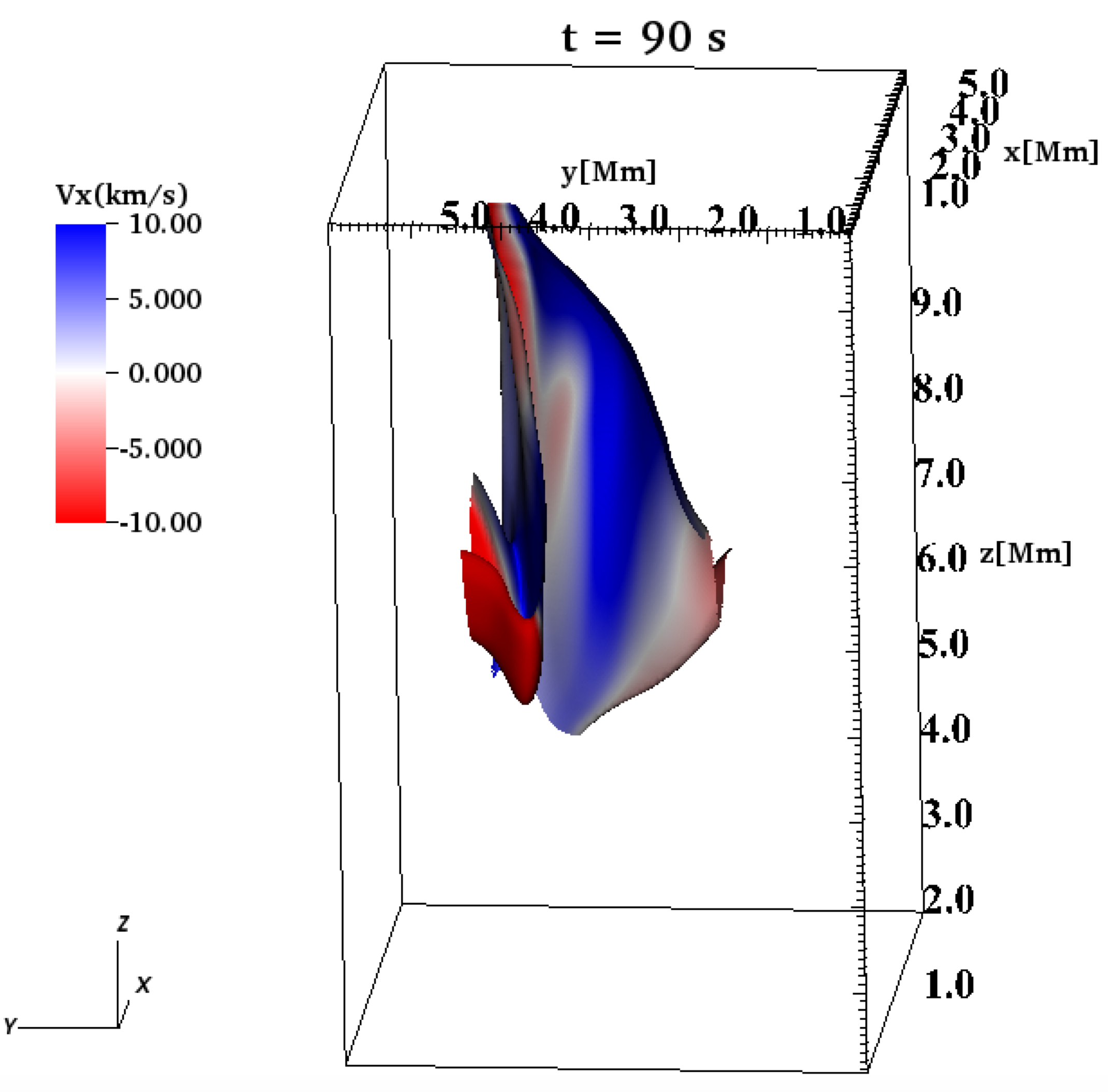}
\includegraphics[width=5.5cm]{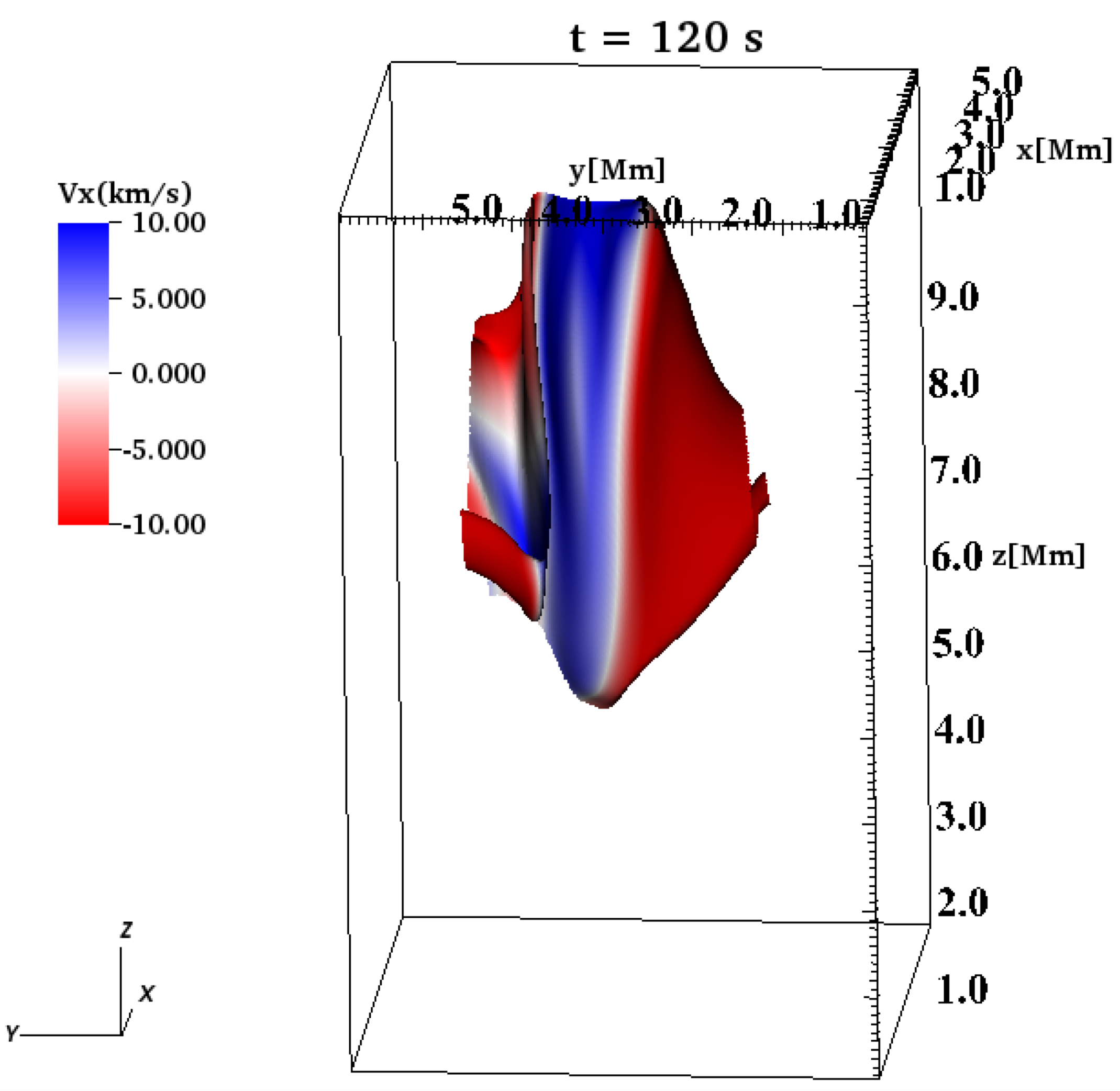}\\
\includegraphics[width=5.5cm]{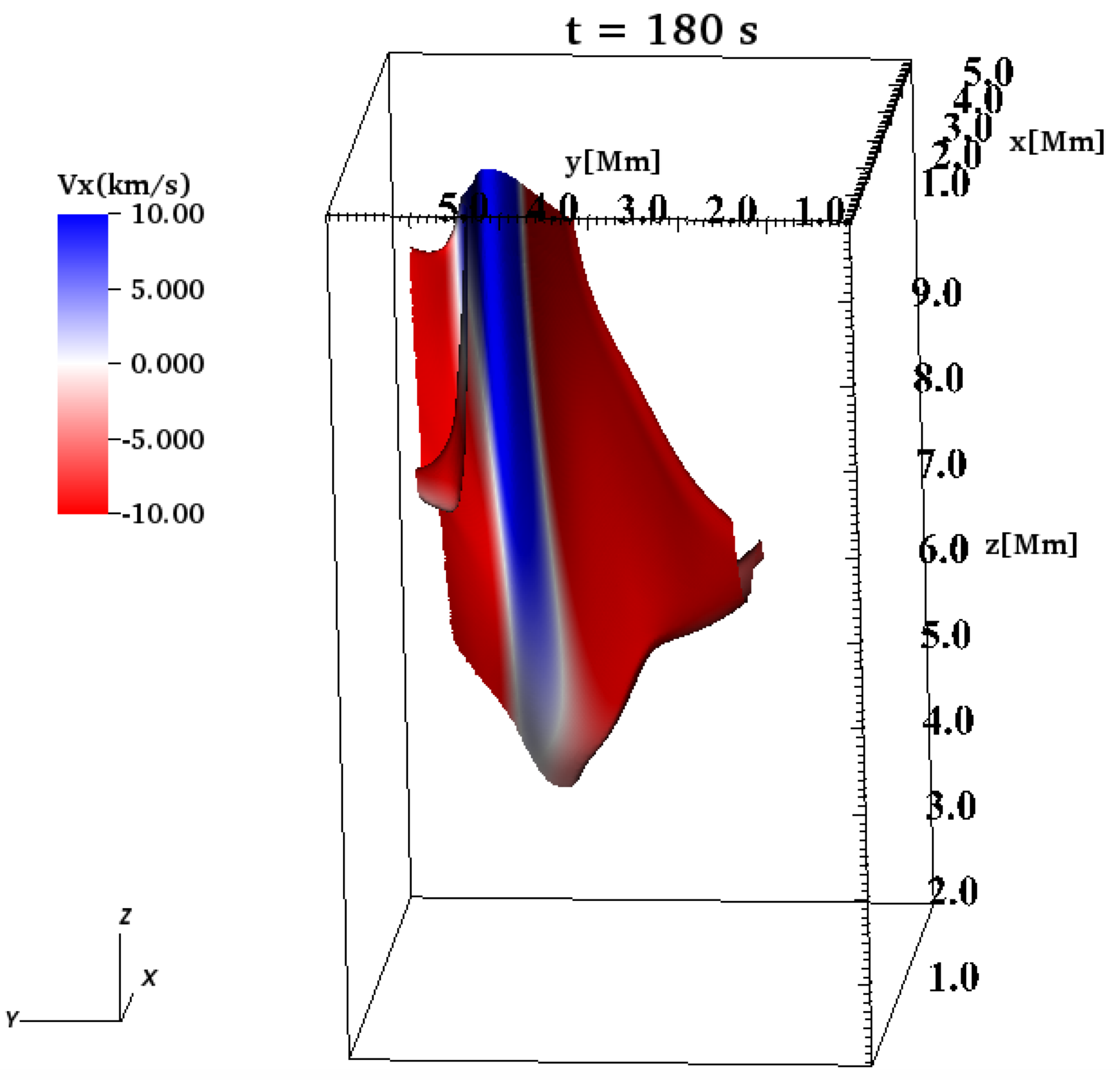}
\includegraphics[width=5.5cm]{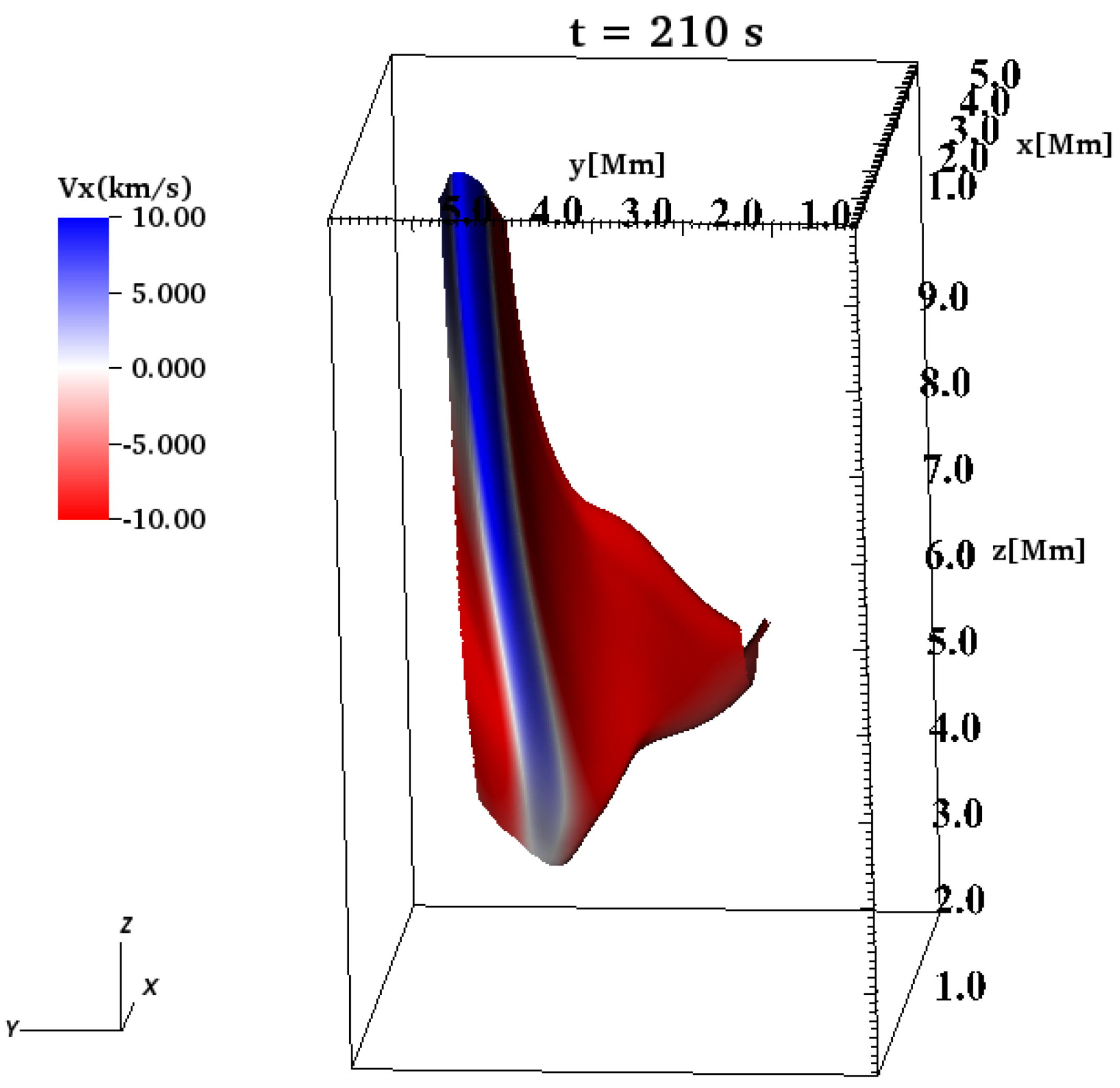}
\caption{\label{3D_temp_contours_vx_color_maps} Snapshots of an isosurface of the plasma temperature equal to $10^{4}$ K at various times. The colours of the isosurface represent the value of $v_{x}$ in km s$^{-1}$. Courtesy of \citet{Gonzalez-Aviles_et_al_2019}, Figure 4.} 
\end{figure*}

\begin{figure*}
\centering
\includegraphics[width=0.99\textwidth]{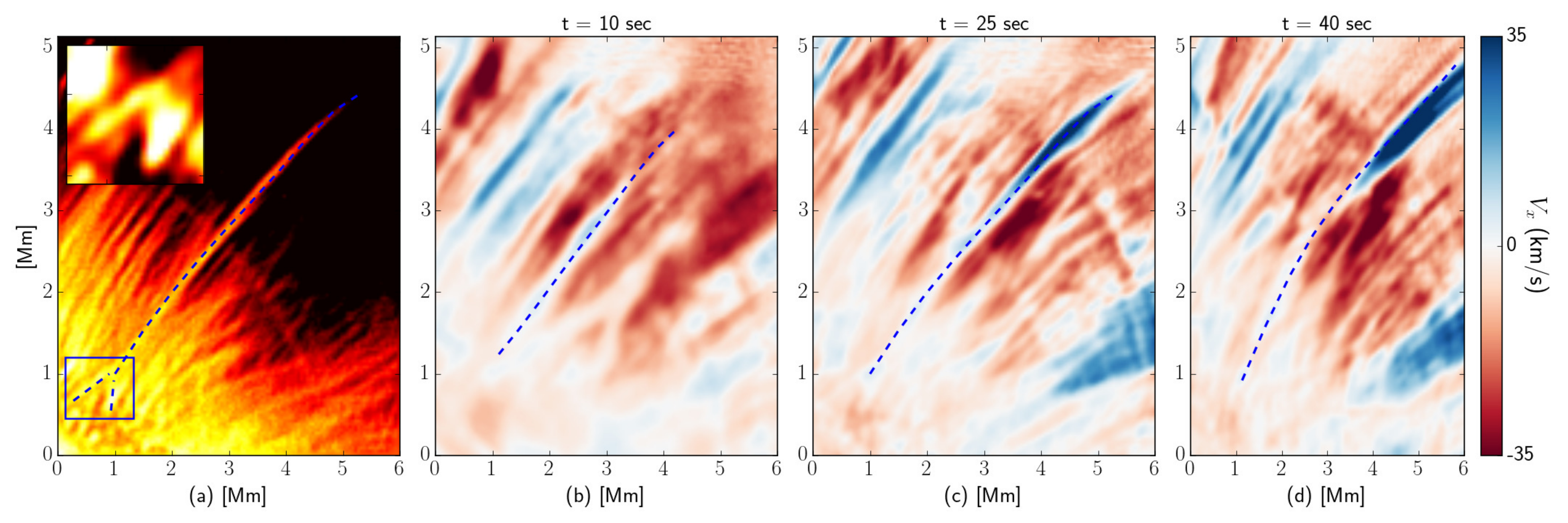}
\caption{\label{doppler_spicule} From left to right, we show an intensity image of an observed spicule off-limb in H$\alpha$ and the evolution of the LOS Doppler velocity for three times, see Figure 5 in \cite{Gonzalez-Aviles_et_al_2019}.}
\end{figure*}
Numerical simulations of jets also demonstrate wave behaviour throughout their evolution. For example, the 3D numerical modelling performed by \citet{Gonzalez-Aviles_et_al_2018} shows that a jet with attributes of a type II spicule exhibited rotational and torsional motions, suggesting the presence of Alfv\'en waves excited directly in the corona. This result is significant as it implies that Alfv\'en waves excited by jet evolution do not need to propagate through the lower layers of the solar atmosphere; instead, they may deposit their energy directly in the corona.

Furthermore, the study by \citet{Gonzalez-Aviles_et_al_2019}, previously discussed in Section \ref{formation}, also investigated the role of MHD waves in a 3D simulated solar jet. An example of the evolution of a jet's dynamics is shown in Figure~\ref{3D_temp_contours_vx_color_maps}, where the contours of temperature with a value of 10$^{4}$ K are colored with the distribution of $v_{x}$ at various times. The behavior of the jet (see Figure~\ref{3D_temp_contours_vx_color_maps}) is similar to the observed spicules seen off-limb in the H$\alpha$ wavelength displayed in Figure~\ref{doppler_spicule}. In the first panel of Figure~\ref{doppler_spicule} the intensity map of a spicule shows an inverted $Y$-shaped structure at the footpoint, suggesting the presence of a magnetic reconnection process, similar to the one proposed by, e.g., \cite{Shibata_et_al_2007} and \cite{ He_et_al_2009}. The next three panels of Figure~\ref{doppler_spicule} show the temporal evolution of LOS Doppler velocity at three different times. At the time $t=25$ s, the observed spicule shows strong blue-shift Doppler velocity, which is similar to the simulated jet's behavior at $t=120$ s, shown in Figure
~\ref{3D_temp_contours_vx_color_maps}. Finally, at time $t=40$ s, the top of the spicule shows asymmetry in the LOS Doppler velocity, indicating rotational motion. This behavior is similar to the results obtained in the simulated jet shown in Figure~\ref{3D_temp_contours_vx_color_maps} at $t=180$ s. Apart from the rotational motions, the simulated jet shows a kink oscillation (see Figure 3 in \cite{Gonzalez-Aviles_et_al_2019}), which represents a superposition of coupled rotational and transverse motions \citep{Goossens2014}. In the corona, the rotation in a jet can be generated by (i) kink motion or (ii) by the Lorentz force. In fact, for the lower solar atmosphere where magnetic forces are less dominant, the study by  \citet{shel2011} showed that vorticity can be generated even in the absence of a magnetic field in the solar photosphere. Furthermore, recent studies by \citet{Silva2020, Yasir2022} showed that hydrodynamic forces are comparable (in strength) with Lorentz forces inside vortex structures, even at the chromospheric level.

By combining observations of type II spicules with numerical data, the study by \cite{Chintzoglou2018ApJ} suggested that the high propagating speeds observed in network jets (disk counterparts of spicules when observed using IRIS) may not be due to the plasma motion. Instead, based on numerical evidence, the authors suggest that the speed can be attributed to a rapidly propagating heating front generated by the diffusion of electrical currents from ambipolar diffusion propagates along an already pre-existing spicular structure. Therefore, this heating front may be misinterpreted as mass up-flow motion when, in reality, it provides the observed upward speeds of network jets.

\section{Summary and future work}\label{summary}
Jets in the lower solar atmosphere could considerably contribute to the energy budget of the solar atmosphere and provide a plausible conduit of mass, momentum, and energy transfer through the layers of the solar atmosphere. This review summarizes the current state-of-the-art understanding of jet dynamics in the solar atmosphere, including formation and evolution from an observational and numerical perspective. In particular, we discussed possible physical mechanisms responsible for jet formation and the types of MHD waves that such jets can excite. Numerical simulations show jets that exhibit torsional motion, which, in observations, would manifest as red-blue axisymmetry in LOS Doppler velocity across the width of the jet. Furthermore, numerical simulations have also shown that torsional motions of jets may appear at coronal heights due to the Lorentz force acting on the plasma. Thus, the jet's rotation could explain the more pronounced red-blue Doppler velocity axisymmetry in spicules at coronal heights observed with the CRISP instrument (shown in Figure~\ref{doppler_spicule}).

Recent advances in resistive MHD numerical modeling aimed to explain the physics behind jet formation in terms of magnetic reconnection. Examples of jets generated in 2D and 3D simulations had a striking resemblance to type II spicules' morphology, velocities, and lifetimes. In addition, numerical simulations have also hinted at the evidence of plasma instabilities, which may be a catalyst for energy cascade to smaller spatial scales, which is necessary to generate turbulence and local plasma heating.

While this review focuses on small-scale solar jets typically found in the lower solar atmosphere, larger-scale jets such as surges, macro-spicules, and coronal jets release energy at much greater spatial scales, are also of interest. These larger-scale solar jets may share some similar dynamical characteristics with chromospheric jets. For further details about the observations, theory, and modelling of large coronal jets, see the review by \citet[][]{Raouafi2016, shen2021}.

Since ubiquitous small-scale jets (e.g., spicules) have a very narrow width relative to their length, detection of instabilities forming at their edges has not been possible due to the inadequate spatial resolution of current instruments. That is why forthcoming observations by DKIST, the largest solar telescope to date, will provide a much-needed leap forward in understanding instabilities and waves in solar jets. Furthermore, future missions, such as SULIS, will help reconstruct the 3D coronal magnetic field that is an essential parameter for detecting how the magnetic field evolves along the edges of larger-scale coronal jets.

\subsection{Future objectives}
In recent decades, significant progress has been made with the ground, and space-based telescopes, such as SDO/AIA, Hinode/SOT, SST, IRIS, have provided more detailed observations of jets and accelerated research in this field. In addition, observations from these instruments both individually and collectively have given researchers greater insight into the structure and dynamics of small-scale features, including jets observed in the chromosphere/interface region. Below are some highlighted targets for future observational missions and further work to advance theoretical and numerical understanding.

\subsubsection{Increased spatial and temporal resolution observations}
Despite the tremendous progress made over recent decades, observations with a much greater spatial and temporal resolution are required to aid understanding of the small-scale physics describing the dynamics of such jets. Fortunately, the new high-resolution observational facility DKIST will provide new important information on small-scale solar jets' finer structure and dynamics. DKIST is capable of capturing the dynamic evolution of jets, such as spicules, with high cadence (3 seconds) and making simultaneous measurements (with 15-second cadence) of both chromospheric and photospheric magnetic fields and flows that underlie jet initiation and acceleration \citep{Rast2021SoPh}. Together these will help our understanding of the formation mechanisms behind small-scale solar jets and the roles that other small-scale processes such as magnetic reconnection play in their formation. In addition, DKIST will hopefully provide a deeper understanding of the mass and energy cycle in the lower solar atmosphere and answer outstanding questions such as the role of spicular type jets in the energy balance of the solar corona?

\subsubsection{Partially ionised plasma}
At various points in this review, discussions have arisen relating to previous studies focusing on partially ionised plasma and its role in, e.g., the formation of spicular jets. Compared to the hot, tenuous corona, the lower solar atmosphere is relatively dense and weakly ionized. Ion-neutral interactions can lead to increased resistivity and an increase in the effective ion mass, with a consequent reduction in the Alfv\'{e}n speed. As a result, the plasma is no longer in the framework of ideal MHD, and these factors have significant consequences for solar phenomena. The effect of collisions between neutrals and ions likely results in the damping rate of MHD waves is increased, affecting the Alfv\'{e}n wave energy flux into the corona and the behaviour of, e.g., MHD shock waves being altered. However, there is much still unknown about the true effects a multi-fluid plasma has on jets' formation and evolutionary dynamics, so observational evidence and verification of partial ionisation effects are vital in determining its role in the properties of small scale solar jets.

\subsubsection{Wave analysis}
MHD waves can have considerable implications in the energy balance of the solar atmosphere. Not only are these waves ubiquitous throughout the solar atmosphere, but as discussed earlier, they are commonly observed propagating within solar jets. However, critical questions regarding MHD waves in the solar atmosphere remain unanswered, such as what processes drive these waves? What is the effect of non-uniform plasma and flow on the characteristics and propagation of these waves? Can solar jets support mode conversion of MHD waves? How does the energy contained within solar jets propagate into the upper layers of the solar atmosphere, and eventually, what are the physical mechanisms which can dissipate this energy in the solar corona? In order to answer these questions by studying small-scale jets, it is essential to have observations of phase relations of oscillations at different heights of the jet in the solar atmosphere complemented with information about neighbouring jets in the local environment. Observations relating to these aspects will help us understand the nature of waves within jets, e.g., whether they oscillate due to the same underlying global motions, whether they are propagating or standing, and the physical properties of the waves. All these are crucial factors for conducting jet seismology.

\subsubsection{Analytical modelling}
For accurate seismology to be conducted using observations
of  MHD waves, it is necessary to have a realistic analytical model of jets as a waveguide to combine this with high-resolution observations. The typical representation of a jet as a uniform magnetic cylinder with a background plasma flow is convenient in the analytic sense; however, in practice comes with many limitations. The lower solar atmosphere, which, as discussed, is replete with small-scale plasma jets, must be modelled with waveguides that have non-uniformity of plasma and background flow \citep{Skirvin2021a, Skirvin2022} and a partially ionised plasma in which it is embedded. Furthermore, realistic jet models that vary with altitude/jet height are required as small-scale solar jets frequently propagate between regions of spatially varying plasma-$\beta$, which could dramatically affect the resulting wave modes and their properties. While these models are complex, a numerical approach may complement advances in analytical knowledge.

\subsubsection{Numerical simulations}
Even though there have been significant advances in numerical simulations on jets and their associated phenomena, further progress must still be made. For instance, numerical models should address the fine scales and compare more precisely with observations. Thus, there must be improvements to numerical codes to exploit the more recently available computational resources, such as Message Passage Interface (MPI) and the Compute Unified Device Architecture (CUDA). Furthermore, the simulations must include all the possible physical properties of the solar atmosphere, for example, the partial ionization coupled with the full radiative transfer equations, which could better describe the conditions of all the layers covering from the convection zone to the solar corona. Current methods to model chromospheric radiative transfer and non-equilibrium ionisation are the least precise techniques in comparison to those applied to describe other parts of the solar atmosphere \citep{Leenaarts_2020}. Additionally, numerical simulations must be improved to model scenarios where the plasma dynamics are more realistically highly non-linear; therefore, we ought to exploit and improve their capabilities and continue to analyse the complexity of the jets and their associated phenomena.

%\acknowledgments    %SOLAR PHYSICS

\section{Acknowledgments}    %ASR
V.F., G.V., and I.B. are grateful to The Royal Society (International Exchanges Scheme, collaboration with Mexico, Chile, and Brazil) and Science and Technology Facilities Council (STFC) grant no ST/V000977/1 for the support provided. This work also greatly benefited from the discussions at the ISSI workshops `Toward Dynamic Solar Atmospheric Magneto-Seismology with New Generation Instrumentation' and `The nature and physics of vortex flows in solar plasmas'. S.J.S. would like to thank STFC (UK) for the PhD studentship project reference (2135820). F.G. and J.J.G.A. thank Royal Society-Newton Mobility Grant NI160149, CIC-UMSNH 4.9, and CONACYT 258726 (Fondo Sectorial de Investigaci\'on para la Educaci\'on). J.J.G.A also thanks ``Investigadores por M\'exico-CONACYT" (CONACYT Fellow), CONACYT LN 315829 (proyecto 2021) and CONACYT-AEM Grant 2017-01-292684 for partially support this work. The program "Investigadores por M\'exico-CONACYT", project 1045 sponsors the Space Weather Service Mexico (SCiESMEX). This research has also received financial support from the European Union's Horizon 2020 research and innovation program under grant agreement No. 824135 (SOLARNET). The simulations by \cite{Gonzalez-Aviles_et_al_2017,Gonzalez-Aviles_et_al_2018,Gonzalez-Aviles_et_al_2019} were carried out in the facilities of ``Centro de Superc\'omputo de  Clima Espacial (CESCOM)" part of the ``Laboratorio Nacional de Clima Espacial (LANCE)" and the Big Mamma cluster at the LIASC-IFM. Figures 6, 7 and 12 were generated with the use of the VisIt software package \citep{Childs_et_al_2012}. Parts of this research have been undertaken with the assistance of resources and services from the National Computational Infrastructure (NCI), which is supported by the Australian Government. Some of the results were obtained using the OzSTAR national facility at Swinburne University of Technology. The OzSTAR program receives funding in part from the Astronomy National Collaborative Research Infrastructure Strategy (NCRIS) allocation provided by the Australian Government. This research was supported partially by the Australian Government through the Australian Research Council's Discovery Projects funding scheme (project DP160100746) and through Future Fellowship FT120100057 awarded to Dr Sergiy Shelyag. The views expressed herein are those of the authors and are not necessarily those of the Australian Government or Australian Research Council.

%%%%    FOR SOLAR PHYSICS   %%%%%

     % format of references provided by the journal (.bst)
%\bibliographystyle{spr-mp-sola}
     % name your Bibtex file containing your references (.bib)
%\bibliography{agu} 

%%%%    FOR ASR   %%%%%
\bibliographystyle{model5-names}
\biboptions{authoryear}
\bibliography{asr}

%\end{article} 

\end{document}